# Upconverting microgauges reveal intraluminal force dynamics *in vivo*

Jason R. Casar[7*], Claire A. McLellan[7], Cindy Shi[7], Ariel Stiber[7], Alice Lay[1], Chris Siefe[7], Abhinav Parakh[2,7], Malaya Gaerlan[3,7], Wendy Gu[4], Miriam B. Goodman[5*], Jennifer A. Dionne[6,7,8*]

[1]Department of Applied Physics, Stanford University, Stanford CA, 94305, USA

[2]Materials Engineering Division, Lawrence Livermore National Laboratory, Livermore CA, 94550, USA

[3]Department of Biology, Stanford University, Stanford CA, 94305, USA

[4]Department of Mechanical Engineering, Stanford University, Stanford CA, 94305, USA

[5]Department of Molecular and Cellular Physiology, Stanford University, Stanford CA, 94305, USA

[6]Department of Radiology, Stanford University, Stanford CA, 94305, USA

[7]Department of Materials Science and Engineering, Stanford University, Stanford CA, 94305, USA

[8]Chan Zuckerberg Biohub, San Francisco, San Francisco, CA 94158

*Address correspondence to: jcasar@stanford.edu; mbgoodmn@stanford.edu; jdionne@stanford.edu

**Abstract**

The forces generated by action potentials in muscle cells shuttle blood, food, and waste products throughout the body's luminal structures. While non-invasive electrophysiological techniques exist,[1–3] most mechanosensitive tools cannot access luminal structures non-invasively.[4–6] Here, we create non-toxic, ingestible mechanosensors to enable the quantitative study of luminal forces and apply them to study feeding in living *Caenorhabditis elegans* roundworms. These optical "microgauges" comprise upconverting $NaY_{0.8}Yb_{0.18}Er_{0.02}F_4@NaYF_4$ nanoparticles (UCNPs) embedded in polystyrene microspheres. Combining optical microscopy and atomic force microscopy to study microgauges *in vitro*, we show that force evokes a linear and hysteresis-free change in the ratio of emitted red to green light. With fluorescence imaging and non-invasive electrophysiology, we show that adult *C. elegans* generate bite forces during feeding on the order of 10 µN and that the temporal pattern of force generation is aligned with muscle activity in the feeding organ. Moreover, the bite force we measure corresponds to Hertzian contact stresses within the pressure range used to lyse the worm's bacterial food.[7,8] Microgauges have the potential to enable quantitative studies that investigate how neuromuscular stresses are affected by aging, genetic mutations, and drug treatments in this and other luminal organs.

**Main**

Hollow neuromuscular organs, such as those of the cardiovascular and gastrointestinal systems, rely on electrical induction of muscle activity to generate force and propel material. Dysfunction in the magnitude, frequency, or coordination of force generation is a hallmark of motility diseases. For example, contractile pathologies can affect bolus transport in the esophagus,[9] bladder function in patients with nervous system damage,[10] and heart rhythms in patients with inherited arrhythmia syndromes.[11] Unfortunately, a quantitative understanding of the local pressure gradients generated by electrical signaling remains incomplete due to a lack of miniaturized mechanosensing devices to access neuromuscular cavities non-invasively. The study of force in biology has been advanced by techniques such as optical tweezers, atomic force microscopy (AFM), and traction force microscopy (TFM), but these instruments cannot access *in vivo* tissues.[4,12] Catheter-based techniques such as manometry[13,14] are limited to larger organs, and while micromachined mechanosensors show promise,[6] their use requires surgical implantation. Fluorescent nanomaterials, including Förster resonance energy transfer-based strain sensors,[15] can overcome some constraints imposed by luminal organs due to their small size and optical readout modality. Indeed, such sensors have led to remarkable insights into mechanosensation from the cellular[16,17] to the organism level.[18] Their dynamic range (~1-10 pN) generally limits their utility to stresses induced on or by proteins.[5] Moreover, the propensity of fluorophores to rapidly photobleach makes tracking force actuation over minutes a difficult task, and their excitation by visible light creates strong autofluorescence, decreasing their signal-to-noise ratio. To date, the field of mechanobiology lacks techniques that are well suited for measuring the relatively large compressive forces collectively actuated by many muscle cells inside the internal, narrow, and tortuous structures of the body's tubular cavities.

Mechanosensitive upconverting nanoparticles (UCNPs) are a class of biocompatible[19–21] materials that have been explored for next-generation optogenetics[22] and phototherapy[23] applications. Their popularity in biological applications stems from their ability to generate visible emission from low-energy infrared excitation, which induces negligible autofluorescence background and penetrates deeply into tissues.[24] The distinct anti-Stokes shift is possible via the sequential non-radiative energy transfer (ET) between long-lived 4f states of trivalent lanthanide dopant pairs.[25] Our group has previously demonstrated the ratiometric sensitivity of red ($^4F_{9/2} \rightarrow {}^4I_{15/2}$) and green ($^2H_{11/2} + {}^4S_{3/2} \rightarrow {}^4I_{15/2}$) emission lines (Fig. 1a) within core@shell cubic-phase ($\alpha$) NaY$_{0.8}$Yb$_{0.18}$Er$_{0.02}$F$_4$@NaYF$_4$ UCNPs to (quasi)hydrostatic pressure.[26–28] This material exhibits an optimal balance between high sensitivity and brightness, undergoing a visible color change that can be registered on a standard RGB camera.[27]

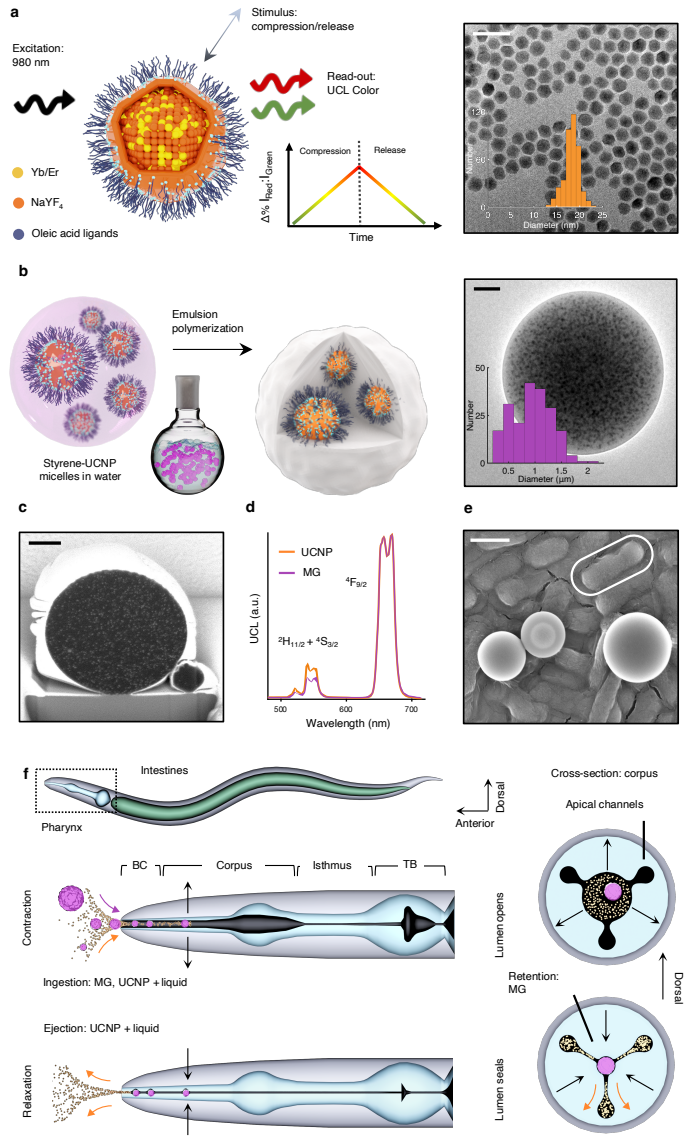

**Figure 1** Synthesis of biocompatible microgauges compatible with passive ingestion

**a)** Left: A schematic of a core@shell $\alpha$-NaY$_{0.8}$Yb$_{0.18}$Er$_{0.02}$F$_4$@NaYF$_4$ UCNP mechanosensor and its operation, in which UCL color ($\Delta\%$I$_{Red}$:I$_{Green}$) becomes redder during compression. Right: TEM image of core@shell UCNPs and a histogram of their diameters (18.2 ± 1.7 nm; mean ± s.d., $n$ = 395). **b)** Left: A schematic of the emulsion polymerization scheme, wherein UCNP/styrene micelles undergo polymerization once heated. Right: A TEM image of a single microgauge with UCNPs visible and a histogram of microgauge diameters (935 ± 367 nm, $n$ = 202). **c)** An SEM image of a single microgauge cross section after ion beam milling (see Supplementary Video 1 for all cross sections). Platinum deposition (outer white), polystyrene (black), and UCNPs (white flecks) are visible. **d)** $\alpha$-NaY$_{0.8}$Yb$_{0.18}$Er$_{0.02}$F$_4$@NaYF$_4$ UCL spectra (exc. 980 nm) before (orange) and after (purple) polystyrene encapsulation. **e)** An SEM image of microgauges on *E. coli* (outlined in white). **f)** A schematic of a *C. elegans* roundworm. Below is a magnified pharynx undergoing the contraction (top) and relaxation (bottom) phases of a pharyngeal pump. Right: procorpus cross sections showing the central lumen contracting open (top) and relaxing closed (bottom). The black arrows indicate the direction of muscle movement, and the orange and purple arrows indicate the direction of UCNP and microgauge movement, respectively. BC = buccal cavity, TB = terminal bulb. Scale bars are 50 nm (**a**), 200 nm (**b**), 500 nm (**c**), and 1 μm (**e**).

We develop $\alpha$-NaYF$_4$ UCNP-embedded polystyrene microbeads (referred to as "microgauges" below) that are sensitive to micronewtons of uniaxial compressive force, and we calibrate their mechano-optical response using dual confocal optical and atomic force microscopy (confocal AFM). To demonstrate their utility, we use the *C. elegans* nematode pharynx, a luminal neuromuscular organ responsible for the mechanical digestion of bacterial food. The action and mechanisms of pharyngeal pumping, the animal's method of feeding, bear striking similarities to those of the human heart, including homologous ion channels that regulate action potentials.[29–31] Combined with the ease of *C. elegans* cultivation and fluorescence imaging, these factors make pharyngeal pumping an attractive model behavior for sensor demonstration.[32] The polystyrene vessel mimics the size of the worm's bacterial food source, allowing it to bypass a size-selective filtering mechanism after ingestion, and provides a controlled ET environment during luminal transit. Ingested microgauges exhibit no toxicity and respond to increasing compressive forces with an increase in their red-to-green emission ratio, which we record in concert with the action potentials that regulate feeding. We observe a temporal correlation between the electrical induction of contraction and the ratiometric emission profile in the pharyngeal lumen of live *C. elegans* roundworms and find that this tissue generates an average maximal force increase on the order of 10μN. This system utilizes correlated electrophysiology and mechanical imaging in live organisms to directly measure bite force in *C. elegans*. Microgauges offer a framework for linking genetic disorder to defects in force generation in neuromuscular organs. However, their future application is not limited to *C. elegans*, as microgauges can be customized to other nematodes and other animals.

Core@shell $\alpha$-NaY$_{0.8}$Yb$_{0.18}$Er$_{0.02}$F$_4$@NaYF$_4$ UCNPs are synthesized via thermal decomposition and hot injection.[28] Trivalent ytterbium and erbium dopants in the nanoparticle cores cooperatively convert incident 980 nm near-infrared excitation into visible emission with two prominent lines in the red

(660 nm) and green (520 + 545 nm) spectral regions (Fig. 1a,d). The inert $NaYF_4$ shell decreases the ET rate from dopants in the core to vibrational modes on the solvent, enhancing the UCL intensity.[33] Brightness is further enhanced once the UCNPs are packaged in hydrophobic polystyrene, especially when delivered to the water-filled pharynx. Aqueous environments pose a significant challenge to UCNP sensor performance for several reasons: water can rapidly disintegrate the lattice,[34] its high-energy -OH vibrational modes efficiently quench UCL,[35] and aqueous solutes can affect the emission color, brightness, and lifetime in ways that might compete with mechanical stimuli.[36] We avoid this signal loss and stimulus competition by packaging UCNPs in a controlled chemical environment via a modified emulsion polymerization method,[37] as depicted in Fig. 1b. This procedure results in UCNPs that are randomly and densely distributed throughout the interior of the polymer host, as revealed by transmission electron microscopy (TEM) images of whole microgauges (Fig. 1b) and scanning electron microscopy (SEM) images of ion-milled microgauge cross sections (Fig. 1c, SI Video 1). The peak energies of the microgauge UCL spectra are identical to those of the constituent UCNPs, but there is a small decrease in the relative proportion of green photon emission (Fig. 1d). This is likely due to the presence of polystyrene, the aromatic -CH stretching modes of which have better resonance with the green quenching $^4S_{3/2} \rightarrow {}^4F_{9/2}$ transition than the red quenching $^4F_{9/2} \rightarrow {}^4I_{9/2}$ transition (ED Fig. 4). Overall, this encapsulation procedure affords UCNPs a brightness that is consistent and sufficient to image with high temporal resolution (20 ms), as detailed in later sections.

In addition to maximizing brightness in the aqueous environment of the lumen, this encapsulation allows UCNPs to bypass the size-selective filtering mechanism of the pharynx, which will eject material smaller than ~200 nm.[38] Like the human heart, the pharynx is a semi-autonomous neuromuscular organ, which rhythmically contracts and relaxes in response to action potentials.[30] In the corpus and anterior isthmus, the outward radial contraction of epithelial muscles opens the pharyngeal lumen, drawing particle-laden fluid in through the buccal cavity (Fig. 1f). The subsequent relaxation of these muscles seals the lumen, retaining large particles at trap sites in the anterior corpus (procorpus) and anterior isthmus while ejecting fluid and smaller particles. The procorpus cross sections in Fig. 1f depict how this is achieved. Small material is diverted radially outward through narrow constrictions between the central lumen and apical channels, where it can bypass the trap sites. Microgauges are too large to be expelled through these constrictions but not so large that they cannot enter through the buccal cavity (~3 μm).[38] Seeking to mimic the primary target of this filtering mechanism, *E. coli* (Fig. 1e), we engineer an average microgauge size of 935 nm (Extended Data Fig. 1). As expected, *C. elegans* allowed to freely ingest Poly(maleic anhydride-alt-1-octadecene) (PMAO)-wrapped UCNPs displayed markedly lower and less frequent UCL intensities than those incubated with microgauge-embedded UCNPs (Extended Data Fig. 2). The pharynx, with its aqueous environment and filtering mechanics, imposes crucial design constraints on UCNP-based mechanosensors, which are efficiently satisfied with the microgauge construct.

**Biocompatibility**

Having constrained the sensor size to mimic the worm's bacterial food source, we can co-opt its natural feeding behavior to passively localize our sensors in the terminal bulb. Figure 2a summarizes the feeding procedure, in which adult animals consume both *E. coli* and microgauges from a small lawn on agar. Microgauge sensors that pass through the pharynx into the intestine are expelled via defecation. There are no obvious differences in animal movement, feeding, or defecation when microgauges are present (Supplementary Videos 3-5). Although their accumulation varies between individuals, microgauges typically accumulate in the anterior isthmus and terminal bulb (Fig. 2a), which is consistent with the current understanding of particle transport in the pharynx.[39]

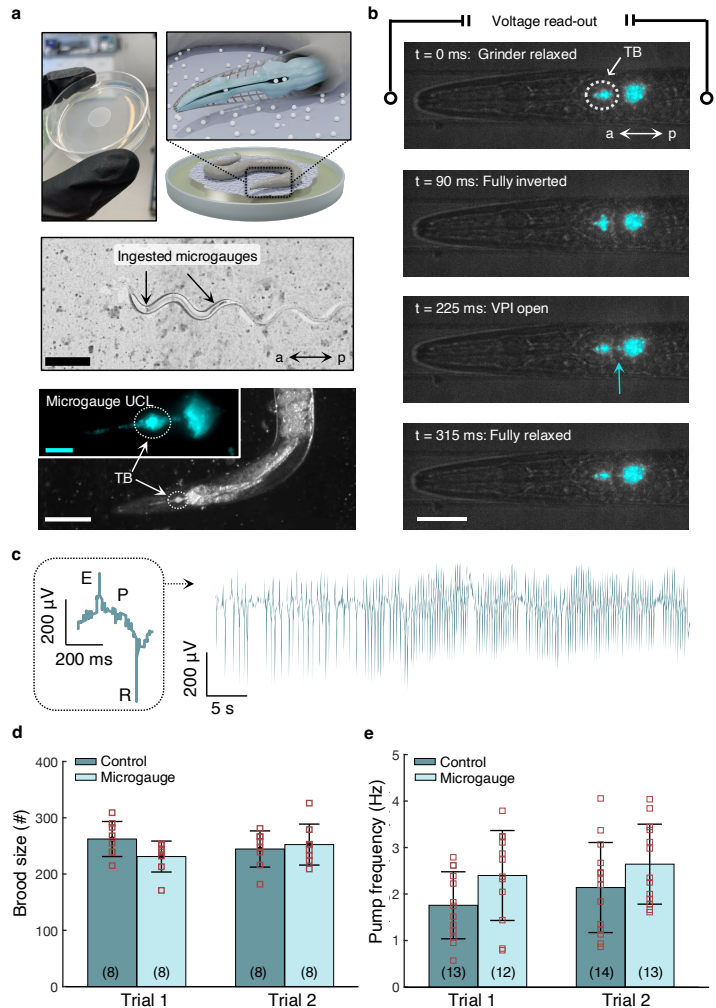

**Figure 2** *Caenorhabditis elegans* ingest biocompatible microgauges

**a)** Top, an agar plate and microgauge lawn, a schematic representation of a worm consuming microgauges from this lawn and a pharynx highlighting the lumen containing ingested microgauges. Middle, an image of a worm on this lawn. Microgauges (dark contrast) are visible in the upper and lower intestines. Bottom, false-colored UCL (cyan) and corresponding brightfield images from ingested microgauges in the pharynx and upper intestines. Scale bars, 250 μm (top), 100 μm (bottom) and 20 μm (inset). **b)** Brightfield and false-colored UCL (cyan) stills of particle transport during serotonin-stimulated pumping for a worm inside the immobilization channel of the EPG chip. Particle transport through the vpi is visible in the third frame and denoted with a blue arrow (source: Supplementary Video 2). Scale bar, 30 μm. **c)** A full 60s EPG time series and waveform (inset) highlighting depolarization (E, contraction onset), plateau (P, contraction maintenance), and repolarization (R, relaxation onset) signal from pharyngeal muscles. **d)** Average brood size (mean ± s.d.) of animals fed *E. coli* with microgauges (light blue; trial 1: 231 ± 28; trial 2: 252 ± 36) or without microgauges (dark blue, trial 1: 262 ± 31; trial 2: 245 ± 32). There was no statistically significant effect of microgauges (trial 1: $p = 0.09$ and trial 2: $p = 0.70$, Student's t-test). Orange circles are results from individual animals, and error bars are standard deviations. **f)** Average pharyngeal pumping rate (mean ± s.d.) from animals fed *E. coli* with microgauges (light blue; trial 1: 2.40 ± 0.97 Hz; trial 2: 2.64 ± 0.86 Hz) or without microgauges (dark blue; trial 1: 1.76 ± 0.72 Hz; trial 2: 2.14 ± 0.97 Hz). There was no significant effect of microgauges (trial 1: $p = 0.09$ and trial 2: $p = 0.17$, Student's t-test). The red squares are the average pharyngeal pumping rates for individual worms over a 60s window, and the error bars are their respective standard deviations. VPI = pharyngeal intestinal valve, a = anterior, p = posterior, TB = terminal bulb.

The grinder resides in the terminal bulb and consists of three radially oriented cells with ridged cuticles. It inverts to mechanically degrade the cell walls of trapped bacterial food.[40] This inversion is nearly synchronous with the onset of contraction in the corpus. Normal transport dynamics during a single representative pump are illustrated in dual UCL (cyan) and bright field video frames (Fig. 2b, Supplementary Video 2). Between frames one and two, the grinder goes from fully relaxed to fully inverted, forcing the microgauges within the terminal bulb posteriorly. Microgauge transport through the partially open pharyngeal intestinal valve (VPI) into the intestines is visible at the blue arrow in the third frame. The pharyngeal muscles then relax, returning the grinder to a resting position. The imaging rate (66 frames per second) is more than 10x the maximum frequency of pharyngeal pumping (~5 Hz). We note that imaging at this rate requires a high irradiance (13 kW/cm$^2$), which induces minor heat stress in the pharynx (Supplementary Fig. 1).

In the first of two biocompatibility assays, we study the chronic toxicity and reproductive effects of microgauges by measuring the total progeny generated during the four-day window from egg-laying onset until cessation (Fig. 2d). In both trials, the fecundity of worms fed a standard diet of *E. coli* are indistinguishable from that of worms fed a diet of *E. coli* mixed with microgauges (Supplementary Fig. 2). Moreover, both the daily number of progeny and the total brood size between the two conditions are similar to those of wild-type worms fed a normal bacterial diet.[41] Similarly, a previous study has found no effect of unencapsulated, ligand-stripped $\alpha$-NaYF$_4$ on worm fecundity,[21] suggesting that UCNPs are non-toxic whether delivered alone or encapsulated in polystyrene.

In the second assay, we measure the effects of microgauge ingestion on pumping frequency, which is estimated non-invasively from an electropharyngeogram (EPG). The EPG reflects the pattern of muscle contractions associated with pumping and can be used to determine their frequency. To measure EPGs, young adult worms were loaded into a specialized microfluidics device and treated with exogenous serotonin to stimulate pumping according to established procedures.[11,42] The inset in Fig. 2c shows the waveform corresponding to a single pump, consisting of an excitation phase (E), followed by a repolarization phase (R). The former triggers the contraction of the anterior pharyngeal muscles - and the terminal bulb muscles with a slight delay - and the latter triggers their subsequent relaxation. Across two trials, the average pharyngeal pumping rate of animals fed a normal bacterial diet is indistinguishable from that of animals fed bacteria plus microgauges (Fig. 2e). Thus, microgauge ingestion and accumulation do not alter pharyngeal function.

**Mechano-optical calibration**

Next, we sought to determine how local, anisotropic compressive loads affect UCL emission color. By integrating laser-scanning confocal microscopy and AFM, performed simultaneous imaging and spectroscopy (Fig. 3a, Supplementary Figs. 4-6). These measurements used anisotropic forces on the order of micronewtons applied across spatial scales that match those of the feeding organ, a choice intended to replicate the likely biological mechanics. The goal of mechanical digestion of bacteria in the terminal bulb is to maximize nutrient retrieval and prevent bacterial colonization of the gut. Although hydrostatic stress may increase slightly as isthmus peristalsis[39] moves material into the sealed compartment of the terminal bulb, the primary stresses used to lyse the bacterial cell walls are anisotropic (compression and shear) and imparted by the grinder. Underlying this assertion is the fact that the guts of worms with defective grinders exhibit much higher live bacterial loads than the wild type.[43–45]

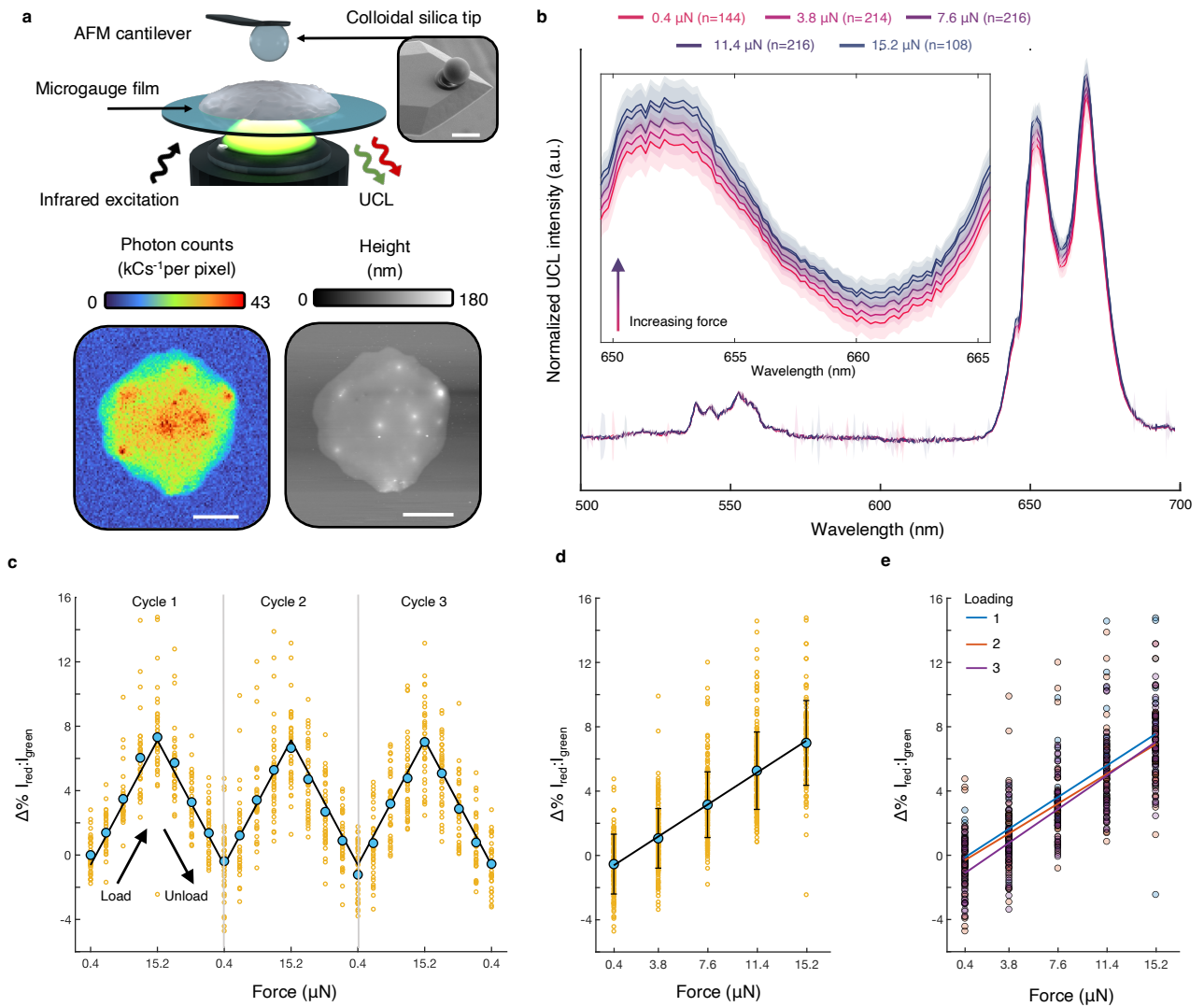

**Figure 3** Optical mechanosensitivity is calibrated using confocal AFM

**a)** Top, a schematic of the calibration setup. An objective below the coverslip-mounted microgauge film simultaneously excites and collects UCL for imaging or spectroscopy. Inset, an SEM image of the indentation tip. Bottom, two coincident images of a sample film: UCL photon counts arriving at a single photon counting module (left), and film height measured in AC mode with a 30 nm radius rounded silicon tip (right). **b)** Averages of 898 green-normalized UCL emission spectra grouped by indentation force. The dark lines represent spectra averages, and lighter regions represent the standard deviation of the wavelength. Inset, an enlarged red peak. **c)** The percent change in the ratio of integrated emission intensities as the film is indented between 0.4 and 15.2 μN in three consecutive cycles. Shown are individual ratio changes from the 12 replicate locations (orange), the averages for each force at a particular order in the loading-unloading cycle (cyan) and the best fit line (black). **d)** The data in **c** grouped by force. **e)** The data in **d**, colored by cycle number, with best-fit lines for each cycle. No statistically significant differences in the slopes of best fit lines are observed at a 95% confidence level. ($p_{1,2} = 0.49$; $p_{1,3} = 0.45$, $n_1 = n_3 = 180$; $n_2 = 179$).

The range of forces used for calibration (0.4 to 15.2 µN) is meant to approximate inactivation stresses *in vivo*. Approximately 1 in 1,000 *E. coli* per hour survive to colonize the intestines,[46] a rate which reflects both the efficacy of the grinder and the ability of unlysed bacteria to adhere to the intestinal wall. Depending on operating conditions, high pressure homogenizers achieve similar lethality rates for *E. coli* between 50 and 200 MPa.[7,8] Informed by these bounds, we employ a Hertzian contact model[47] based on geometrical and material assumptions about *E. coli*,[48] and the cuticular ridges of the grinder[40] to calculate an equivalent uniaxial compressive force in the range of micronewtons to 10s of micronewtons. This conversion also requires knowledge of the compressive modulus of the microgauges (740 MPa), which we derive by fitting 50 single microgauge force indentation curves to a Johnson-Kendall-Roberts (JKR) model (Supplementary Fig. 7).[49] We constrain the upper bound of calibration stress (42 MPa) to less than half of the yield strength of microscale polystyrene (~100 MPa).[50] As expected, we observe no plastic deformation in samples indented under these conditions (Supplementary Fig. 8). Surprisingly, we do not observe permanent divoting in SEM images until indentations exceeding 400 MPa equivalent stress (Supplementary Fig. 9), which could be due to the presence of significantly stiffer ceramic nanoparticles in this composite material. Using a confocal AFM to calibrate local UCL changes against micronewton-scale anisotropic compressive forces, we more faithfully replicate the mechanical conditions that likely exist *in vivo*.

Compressive loads between 0.4 and 15.2 µN result in a linear increase in the ratio of integrated red (634-689 nm) and green (513-566 nm) emission (Fig. 3b). To ensure sample stability during loading-unloading cycles, we sampled thin films formed from microgauge monolayers (Extended Data Fig. 3), and to ensure that the confocal excitation spot collects UCL exclusively from the contact area, we indent with a large 10.2 µm diameter spherical silica tip. Figure 3d shows the change in emission ratio as a function of applied force, which is well fitted by a linear model with slope 0.52% $I_{Red}:I_{Green}$ per micronewtons (s.e.m. = 0.02%). Enhancement of select phonon-mediated (Extended Data Fig. 4) and cross relaxation-mediated (Supplementary Fig. 12) quenching pathways may be responsible for the stress-induced shift toward redder emission. The observed AFM trends are consistent with the linear colorimetric trends reported for various UCNPs under hydrostatic pressure in diamond anvil cell (DAC) studies.[21,26–28] DAC experiments on our microgauges (Extended Data Fig. 5) yield a similar, though dampened, pressure response (5.6% Δ%$I_{Red}$:$I_{Green}$/GPa, vs. 9.2% measured previously for unembedded UCNPs).[28] DAC-measured hydrostatic pressure sensitivity can be converted to an approximate force sensitivity (5.4% Δ%$I_{Red}$:$I_{Green}$/μN) using several simplifying assumptions (Methods). Deviations from these assumptions, as well as the distinction between isotropic stresses obtained in the DAC and anisotropic stress delivered via AFM, may account for the tenfold discrepancy observed.

Confocal AFM measured force responses are mechanically and optically robust. At the average irradiance used in our experiments (400 kW/cm$^2$), variations in power density due to laser output produce a change in $I_{Red}$:$I_{Green}$ of 0.018% (Supplementary Fig. 11), which is below the error in force sensitivity. Responsiveness is independent of UCNP loading (Supplementary Fig. 10). Moreover, the sensitivity is consistent when cycled three times within the elastic regime of polystyrene (Fig. 3c). To verify this, we compare the second and third loadings to the initial loading using two-tailed t-tests (Fig. 3e). No statistically significant differences in the slopes of best-fit lines are observed at a 95% confidence level. Given that replicate measurements are taken sequentially over ~2 hours of continuous laser exposure, the lack of hysteresis also suggests the response is not induced by temperature. Considering the rhythmic pattern of compressions exerted by pharyngeal muscles - and semi-autonomous neuromuscular organs more generally - these data suggest that the mechanical and optical properties of the microgauges will be independent of their loading history.

**Electrophysiology and mechanical imaging**

To demonstrate that these microgauges monitor intraluminal force dynamics *in vivo* and to obtain a quantitative estimate of *C. elegans* bite forces, 60s epochs of correlated two-channel UCL imaging and electrophysiological data were collected from wild-type worms in tandem (Fig. 4a). From these data streams, we obtained traces of the ratio of red and green emission intensities from the terminal bulb of the pharynx (force traces) and of the electrical activity of muscles in the pharynx (EPG), as described (Supplementary Figs. 13-16, Section 7). Individual pharyngeal pumping events are detected in the EPG signal and aligned to the corresponding portion of the force trace (Fig. 4b, Supplementary Fig. 13). As expected from prior studies of pharynx-induced particle movements,[38,39] microgauges are pushed posteriorly during the E-to-R interval, reverse direction after the R phase, and cease moving ~100 ms later (Fig. 4b, Supplementary Video 6).

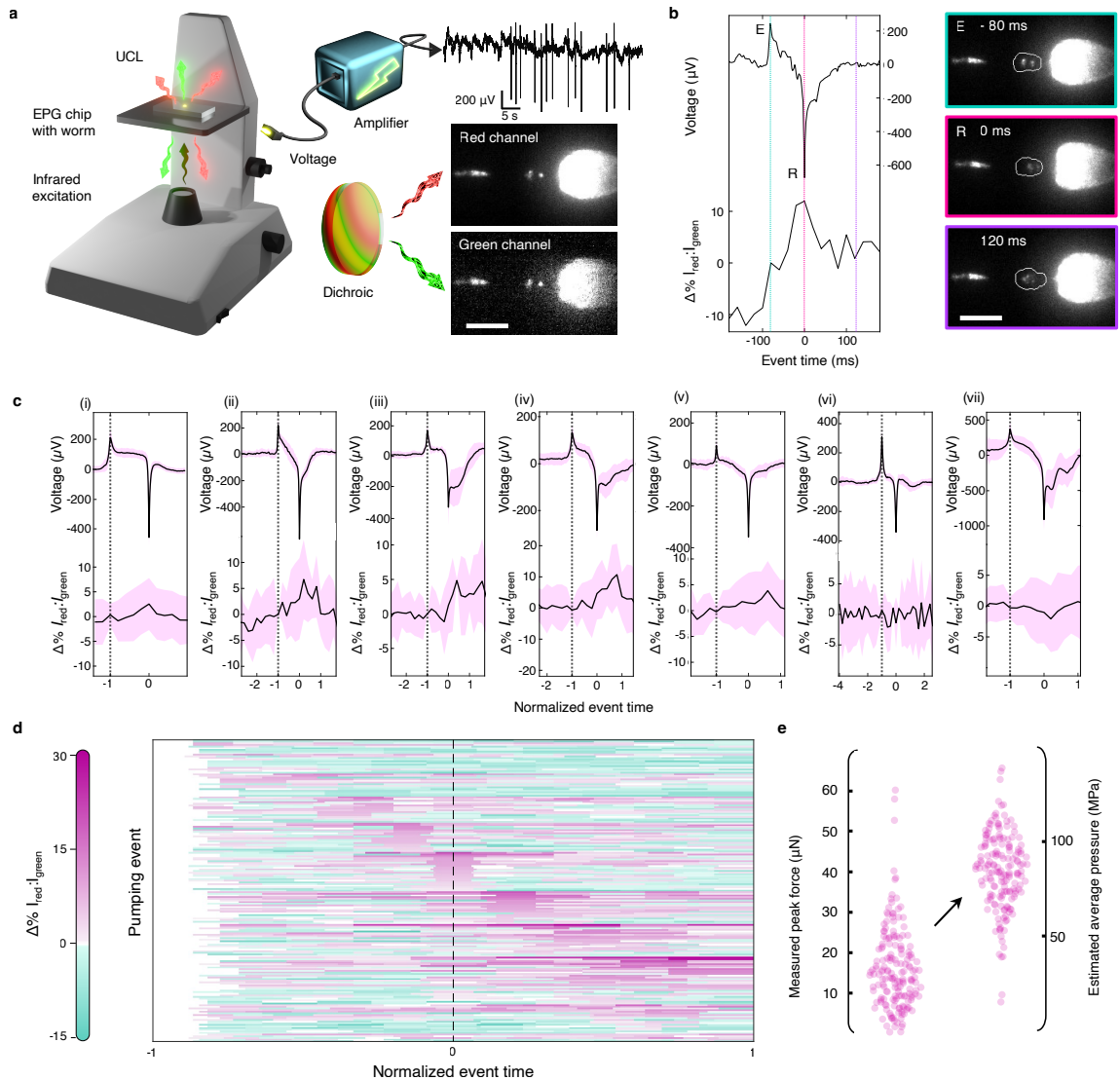

**Figure 4** Microgauges enable simultaneous electrophysiology and optical mechanical imaging of feeding forces in live worms

**a**) A schematic illustrating the correlated electrical and opto-mechanical recording setup, in which UCL luminescence is collected from microgauges within immobilized worms, and then separated into red and green channels (3,000 frames, 50 fps). Example UCL frames and EPG traces correspond to the data set in **c.ii**. Scale bar, 30 μm. **b**) Electrical and optical data for one representative pump, along with the red frames corresponding to the highlighted timepoints. The white outlines surrounding the terminal bulb represent the boundaries of the pixel segments used to extract $\Delta\% I_{red}:I_{green}$. Scale bar, 30 μm. **c**) Event-triggered averages of EPG waveform and $\Delta\% I_{red}:I_{green}$ for all pumps in the 60s recording window, for each of seven worms. Plots i-vii represent each individual animal ($n_i = 70$, $n_{ii} = 13$, $n_{iii} = 12$, $n_{iv} = 18$, $n_v = 29$, $n_{vi} = 18$, $n_{vii} = 25$). Magenta indicates the standard deviation of each normalized time point. **d**) A raster plot of each time-normalized pumping event sorted from top to bottom according to the relative timing of the maximal $\Delta\% I_{red}:I_{green}$ (n = 185). Note that the ratio change at each time point is normalized to the value recorded at E, so all pixels corresponding to E have a value of +0% $I_{red}:I_{green}$ (white). **e**) The maximum force change exhibited over each of the 169 event windows with a detectable force increase (mean ± s.d. = 15.7 ± 10.1 μN), as well as their corresponding average contact pressures (84 ± 20 MPa) converted using a Hertz model assuming a 1 GPa grinder stiffness and an effective contact radius equal to the average microgauge radius, as described in Supplementary Information Section 4.

Each 60s epoch contained dozens of similar pumping cycles, which we leverage to generate event-triggered averages of both EPG and force signals. Because the time separating the E and R phases is variable, it is necessary to align events to the R phase ($t = 0$) and to normalize time to the duration between the E and R phases (see Supplementary Information Section 7). Figure 4c shows event-triggered EPG averages and the corresponding normalized force traces for seven worms. In five worms (i-v), average $\Delta\%I_{Red}:I_{Green}$ values increase within the terminal bulb following contraction. One worm (Fig. 4c.vi) has no detectable change in force and a very short average pump duration (52 vs. the 111 ms average across the other six animals), and another exhibits a modest decrease (Fig. 4c.vii). We note that force traces are derived from the emission of multiple microgauge sensors as they pass through the terminal bulb as the worm feeds, a factor that may contribute to the variation evident in the event-triggered average force traces (Fig. 4c) and across the 185 individual events studied here (Fig. 4d). Very few events (16) lacked a detectable force increase and most (169 of 185) contained a force increase within 100 ms of the R phase of the EPG. In approximately one-third of these cases (60 of 169), the event maximum occurred within 20 ms of the relaxation of the corpus or terminal bulb. This temporal pattern of changes in force is consistent with that of the muscle contractions that govern pharyngeal pumping. Collectively, these findings indicate that microgauges detect the influence of muscle activity on intraluminal stress.

The peak change in the microgauge emission ratio, $\Delta\%I_{Red}:I_{Green}$, is $8.2 \pm 5.2\%$, on average (Fig. 4e). These values are converted into a relative force increase (Fig. 4e) based upon our confocal AFM calibration (Fig. 3d), yielding an average maximum force increase of $15.7 \pm 10.1$ µN. To put this measurement of nematode bite force in context, it is useful to consider that the central function of the terminal bulb and the grinder is to enable nutrient retrieval and to minimize bacterial colonization of the worm's gut. The mean value of average Hertzian contact stress estimated from each peak force ($84 \pm 20$

MPa, as seen in Fig. 4e and described in Supplementary Information Section 4) is consistent with that used in laboratory pressure vessels[8] to achieve a 1,000-fold inactivation of bacterial growth, which is similar to what is observed in living worms.[46] Thus, the stress generated during each pump seems to be sufficient to subserve the prime function of the nematode pharynx.

In summary, we have synthesized, calibrated, and deployed *in vivo* an optically readable sensor of mechanical compression in the lumen of hollow neuromuscular organs. We have demonstrated its viability in quantifying forces associated with pumping in the *C. elegans* pharynx, which is a promising model system for studying arrhythmia and hypercontractility. We envision that this platform could enable the direct comparison of luminal force generation in worms with dysregulation of the duration or frequency of contraction-inducing action potentials in calcium channel mutants[31,39] and how they respond to drug treatment.[29,30] This platform may also find utility as a quantitative, functional replacement for lower-throughput structural assays of muscle mass loss[51] to expedite the study of the effects of caloric restriction, aging, and drug intervention on sarcopenia and healthspan.[52] This proof-of-concept demonstration reveals several opportunities to increase microgauge signals and reduce noise. Utilizing a brighter material with a lower ambient $I_{Red}:I_{Green}$ ratio - such as the "fully-doped" $SrYb_{0.72}Er_{0.28}F_5@SrLuF_5$ - will improve the signal-to-noise ratio of the green emission channel[27] and enable the use of lower excitation power densities to avoid unwanted heating over extended imaging periods. Improved polymerization methods that narrow the distribution of sensor size and UCNP loading density, along with a continuous, in-chip feeding procedure, will help eliminate inconsistencies in absorption and emission cross sections as the worm cycles particles through its terminal bulb. With these improvements, this assay has the potential to directly and non-invasively distinguish intraluminal contractile capability between cohorts of animals and provide insights into the interplay between electrical control and mechanical efficacy.

## Methods

### Upconverting nanoparticle synthesis

The $NaY_{0.8}Yb_{0.18}Er_{0.02}F_4$@$NaYF_4$ UCNPs used to make microgauges were prepared, washed, and stored according to previously reported methods.[28]

### Emulsion polymerization

This procedure was modified from an existing method.[37] Dried UCNPs (15 mg) were resuspended in styrene (1 mL, stabilized, 99%, Acros) and then emulsified in a 5.6 mL aqueous solution of sodium dodecyl sulfate (SDS, 1 mM) and potassium persulfate (KPS, 6 mM) via brief sonication (~2 min). Polymerization of UCNP/styrene micelles into microgauges was achieved after 20 hours of continuous stirring (330 RPM) and heating (66°C) in a vented 20 mL scintillation vial (22G needle). The product was washed in a centrifuge (1000 RCF, 10 minutes, 2x) and allowed to sediment without agitation over a 24-hour period in order to separate the denser microgauges from most of the unembedded or sparsely embedded polymer.

### UCNP surface modification for aqueous dispersal

Aqueous dispersions of single nanoparticles were achieved by wrapping the hydrophobic oleic acid-capped UCNPs with Poly(maleic anhydride-alt-1-octadecene) (PMAO) (Extended Data Fig. 2d). This scheme was modified from an existing method[53] to include 4-dimethylaminopyridine (DMAP, Thermo Fisher Scientific) as an anhydride esterification catalyst, rendering the PMAO more amphiphilic. Dried UCNPs (5 mg) and PMAO (165 mg, Sigma Aldrich Cat. #419117) were first suspended in dichloromethane (DCM, 530 μL) and then combined with MilliQ water (10 mL, Merck-Millipore) and DMAP stock solution (185 μL, 400 mM in DCM). After 30 minutes of sonication, the milky white suspension was transferred to a 50°C oil bath to evaporate the DCM. Once clear and viscous, PMAO-wrapped UCNPs were washed of excess polymer and catalyst in an Amicon-Ultra15 centrifugal filter (100 kDa cutoff) spun at 5,000 RPM (5 min, 3x). The retentate was suspended and stored in 1 mL MilliQ water.

**Microgauge film preparation**

Approximately 10 µg of microgauges were dropcast from suspension onto a plasma hydrophilized silica 0.17 mm thick coverslip (for mechano-optical calibration) or silicon wafer chip (for SEM imaging) and spun at 500 RPM for 30-90s until dry. The resulting microgauge monolayers were placed in a preheated 290°C oven for 30 minutes to allow the polystyrene matrix to melt and then rapidly cooled through the glass transition temperature to room temperature. Most film patches exhibited a gently domed structure that ranged from a minimum of 50 nm thickness at the edges to a maximum of 600 nm thickness at the center.

**Electron microscopy**

A transmission electron microscope (Tecnai G2 F20 X-TWIN, Thermo Fisher Scientific) operating at 200 kV was used to image core and core@shell nanoparticles. Size distributions were measured manually with the open-source image analysis software FIJI.[54] Similarly, size distributions for dropcast microgauges were measured using SEM images taken with a scanning electron microscope (Apreo S LoVac, ThermoFisher Inc.) operating at 5kV and 50 pA. This instrument was also used to image uncoated microgauge monolayers before and after melting and to visualize film surfaces before and after indentation. A focused ion beam scanning electron microscope (FEI Helios Nanolab 600i DualBeam FIB/SEM, Thermo Fisher Scientific) equipped with a Tomahawk $Ga^+$ ion beam was used to assess the 3D distribution of UCNPs in individual microgauge cross sections. In the first step, a conductive platinum layer (500 nm, 51.76 pA) was deposited, and the microgauge was milled enough to expose the interior for focusing. After refocusing, the remainder of the microgauge was alternatively milled (30 kV, 40 pA, 2 µm depth) and imaged (2 kV, 170 pA) with real-time end-point monitoring.

**Confocal atomic force microscopy**

An atomic force microscope head (MFP-3D, Oxford Instruments) fitted with a 10.2 µm diameter spherical silica tip (sQube, CP-NCH-SiO-D5) was mounted on the stage of a microscope (Zeiss Axio Observer) and sealed in an acoustic isolation chamber (Ametek TMC Vibration Control). The spring constant of the cantilever (24.29 nN/nm) was calibrated using the thermal method within the manufacturer's software, and the inverse optical lever sensitivity (155.96 nm/V) was calibrated against a bare silicon coverslip. These values were used to calculate the appropriate voltage trigger for each force. The respective voltage was held for three consecutive 90s spectral acquisitions on a spectrometer (Isoplane SCT-320; 500 nm 300 gr/mm grating) fitted with a digital camera (Blaze 400-LD, Teledyne Instruments). Any spectra affected by cosmic rays within the red or green integration regions were reacquired before advancing to the next voltage trigger. Excitation from a 100 mW 980 nm Obis laser (Coherent Corp.) was focused through a Zeiss 40x 0.95 NA Plan Apochromat objective (400 kW/cm$^2$). Within the collection path of the confocal optical train (Supplementary Fig. 4), two co-aligned single-mode fibers (SM980, 11 mm paf2p-11a collimators, Thorlabs) downstream of a 50:50 beamsplitter cube (CCM1-BS013/M, Thorlabs) acted as pinholes (Supplementary Fig. 5). The "transmission" fiber was used for spectroscopy, and the "reflection" fiber was coupled to a single photon counting module (Excelitas Technologies) for simultaneous imaging. Confocal images were generated by rastering the excitation beam across the sample with a fast-scanning mirror (Newport). Mirror control and fluorescence collection were timed under the control of the imagScan software package (provided by the laboratory of A. Jayich). Contrast differences between the indented (0.4 µN) and unindented microgauge samples were used to place the excitation spot at the center of the tip-sample contact region (Supplementary Fig. 6). To confirm that the excitation power density remained constant throughout the experiment, the axial stability of this coincidence was monitored in real time during the spectral acquisition by monitoring the counts received on the single photon counting module through the second ("reflection") channel, and the lateral stability was confirmed at the end of each cycle by rescanning the confocal image. The percentage change in $I_{Red}$:$I_{Green}$ extracted from these spectra was referenced to the value at the initial 0.4 µN indentation rather than at the unstrained state to account for tip-and cantilever-induced field enhancements.[55] Twelve non-overlapping locations were analysed with this three-cycle ramp.

**Force indentation curves**

A parabolic silicon tip (SD-R30-FM-10, $r$=30 nm, 75 Hz, 2.8 nN/nm) operating in AC mode was used to locate 50 individual microgauges on a glass coverslip. Once located, each microgauge was indented in contact mode up to a 100 nN trigger force at a speed of 1 µm/s, and the force indentation curve was recorded. Indentation is calculated from the difference between the piezo displacement and the cantilever deflection, and force is calculated from the cantilever deflection according to Hooke's law. Baseline corrections of the force-indentation curves were performed automatically within the manufacturer's software. Gradient descent was used to estimate a compressive modulus by fitting the approach curves to the Johnson-Kendall-Roberts model (Supplementary Fig. 7).[49]

**DAC preparation for Raman and UCL measurement**

To prepare the diamond anvil cell, a flake of dried microgauges and silicone oil as a high-pressure medium were loaded into a 300 µm diameter hole drilled using an electrical discharge machining dtool into a T301 stainless steel gasket between two 500 μm diameter diamond culets. Ruby powder was co-loaded as an internal pressure calibrant. After tightening or loosening the set screws, pressure was equilibrated for 30 minutes before acquiring spectra. Both microgauge UCL and ruby photoluminescence were captured on an inverted microscope (Zeiss Axio Observer) fitted with a 20x 0.42 NA 200 mm objective (Mitutoyo), a spectrometer (Princeton Instruments Acton 2500) and a digital camera (Princeton Instruments ProEM eXcelon). Microgauge UCL was excited with a 980 nm diode laser (OptoEngine MDL-III-980, 200 W/cm$^2$), and spectra were acquired with a 500 nm 150 gr/mm grating, a 250 μm wide slit, and a 50s integration time. Ruby photoluminescence was excited with a tunable argon-ion laser at 488 nm (Innova 70c, 900 mW/cm$^2$, Coherent Corp.), and spectra were acquired for 5s with the same spectrometer (Acton 2500; 500 nm 1,800 gr/mm grating, slit width 50 μm). Pressure (in GPa), was calibrated from the fluorescence line shift of the ruby R2 peak according to the relationship

$$\sigma = \frac{1904}{7.715}\left(\frac{\lambda}{\lambda_0}^{7.715} - 1\right) \quad \text{Eq. 1}$$

against an ambient pressure reference wavelength of $\lambda_0$=694.38 nm.[56] The cell was cycled three times consecutively between target minimum and maximum pressures of 0.0001 GPa (ambient) and 6 GPa, respectively. Each cycle consisted of five pressures during loading and one pressure during unloading. Red and green regions of the background-corrected spectra were integrated and the change in their ratio was referenced to the ambient pressure value.

DAC-measured pressure sensitivity was converted into an approximate force sensitivity as outlined previously.[26] Assuming the UCNPs are perfectly spherical with uniform radii equal to the average radius, $r_0 = 9.1\ nm$, experience perfectly isotropic stress, σ, and behave with linear, isotropic elasticity, UCNPs will experience a total force according to

$$F(\sigma) = 4\pi r_0^2 \left(\frac{1}{E^2}\sigma^3 - \frac{2}{E}\sigma^2 + \sigma\right), \quad \text{Eq. 2}$$

where the compressive modulus, E = 272 GPa, was measured previously.[26] We stress that the three assumptions outlined above neglect random error due to UCNP size (coefficient of variation 9%), systematic error in surface area estimates due to non-spherical morphologies, the hydrostatic pressure limit for the silicone oil pressure medium (~3 GPa),[57] and the presence of polystyrene as an effective pressure medium. The presence of polystyrene is particularly consequential, because the solid polystyrene will not propagate hydrostatic stress to its interior, leading to differences between the axial and transverse compression experienced at the UCNP surface, and thus the development of shear stress. Notably, this relationship is also nonlinear. The tenfold difference between confocal-AFM-measured force sensitivity and the force sensitivity approximated from DAC-measured pressure sensitivity via Eq. 2 should be viewed in the context of these simplistic assumptions.

Raman spectra (HR Evolution, Horiba Labram; 600 gr/mm grating; 785 nm excitation) of polystyrene microbeads lacking UCNPs as a function of pressure were taken from inside a DAC (BX-90, One20DAC, Almax EasyLab). Three 60s spectra were acquired and averaged for each pressure. The polystyrene signal overlaps partially with that of the pressure transmitting medium, silicone oil, for which the ambient pressure spectrum is shown in Extended Data Fig. 4b. Because it was challenging to distinguish between the two aliphatic peaks in polystyrene and the lowe- energy aliphatic peak in silicone oil from at all pressures, the three were fitted to a single Gaussian (purple, labeled P-Al+S-Al$_1$ in Extended Data Fig. 4b). The other two peaks correspond to the higher-energy aliphatic peak of silicone oil (red, S-Al$_2$) and the aromatic peak of polystyrene (yellow, P-Ar). Polystyrene microbeads were prepared using the same procedure used to prepare microgauges, but without UCNPs. The wide opening angle (120°) and short working distance (10 mm) simplify optical measurements. The DAC was prepared and operated as described above, with a 150 μm diameter gasket hole and 300 μm diameter diamond culets.

***Caenorhabditis elegans* culturing and microgauge feeding**

Animals were age-synchronized by first isolating eggs from gravid adults via hypochlorite treatment (200 µL household bleach and 20 µL of 5M potassium hydroxide added to gravid adult worms suspended in 1 mL water) and then by transferring eggs to 6 cm growth plates (nematode growth medium (NGM), 2% w/v agar) with food (*E. coli* OP50; optical density at 600 nm, 0.3) and incubated at 20°C. The culturing duration and feeding procedure varied depending on the application. For biocompatibility assays, worms were transferred after incubation for 69 hours to NGM plates seeded with 50 µL *E. coli* OP50, 250 µg SDS, and 100 µg microgauges and allowed to feed for three hours. Separate egg-laying plates with improved optical clarity were prepared in four-well microtiter plates containing 2.5% Gel-rite (RPI) medium ($KPO_4$ (5mM), pH 6, supplemented with $MgCl_2$ (1 mM) and $CaCl_2$, and cholesterol (1:1,000 v/v from stock solution, 5 mg/mL in ethanol)), and each well was subsequently seeded with 50 µL *E. coli* OP50 as described previously (Supplementary Fig. 2).[58] For correlated electrical-optical imaging, worms were transferred at hour 60 to NGM plates seeded with 15 µL *E. coli* OP50 and 100 µg microgauges and allowed to feed for 12 hours. Finally, to compare the ingestion of singly dispersed PMAO-wrapped UCNPs with ingestion of polymer-embedded UCNPs, worms were transferred at hour 69 to feed in M9 buffer without bacteria for one hour under gentle shaking. The buffer consisted of 800 µg microgauges (45 µg UCNPs, see Supplementary Fig. 3 for mass fraction determined with thermogravimetric analysis) or 45 µg PMAO-wrapped UCNPs. M9 (pH ~7) was prepared by mixing 3 g of $KH_2PO_4$ (Sigma-Aldrich), 6 g of $Na_2HPO_4$ (Sigma-Aldrich), 5 g of NaCl (Sigma-Aldrich), and 1L of ultrapure water. For the accumulation comparison, worms were fixed on agarose pads with a drop of 5 mM levamisole hydrochloride (TCI Chemicals) prior to imaging.

**Biocompatibility assays**

To compare progeny among a cohort, two groups of eight worms were cultured and fed as described above, then transferred one at a time to egg-laying plates. Worms were transferred to a new well every 24 hours for a total of four days, and each well was imaged on a flatbed scanner (Perfection v600, Epson), modified as described,[58] 48 hours later. To distinguish debris from mobile worms, reference images of each well were acquired five minutes later. Using FIJI, the progeny present in each well were counter by at least 2 experimenters blinded to each treatment, and the average value was used in each case (Supplementary Fig. 2). To compare pumping rate among a cohort, two groups of 15 worms were cultured and fed as described above.

**Electropharyngeogram (EPG) acquisition**

EPGs were measured using a commercial microfluidics chip (ScreenChip30, InVivo Biosystems). The chip was prepared by flushing it with M9 containing serotonin (4.7 mM, Sigma-Aldrich, Cat. #H9523), to increase the pumping rate, as previously described.[59] Synchronized young adult worms were collected from unseeded NGM growth plates in this buffer and injected into the staging area via a length of polyethylene tubing and a manual syringe. Worms were allowed to adapt to the immobilization channel for 60 s before EPG recording and imaging. The chip was positioned on the stage of an inverted microscope (Zeiss Axio Observer), and the position of the worm in the immobilization channel was verified by observation at low magnification (Zeiss 10x, 0.2 NA EC Epiplan). The applications NemAcquire and NemAnalysis (InVivo Biosystems) were used to collect the time course (500 Hz) and identify waveform features, respectively.

**UCL image acquisition *in vivo***

All UCL from microgauges within the pharyngeal lumen was captured in dual color channels on a digital camera (Orca Flash 4.0, Hamamatsu) fitted with a W-View Gemini Beam Splitter fitted with a 568 nm dichroic (Hamamatsu). Prior to imaging, channel images were aligned by adjusting the internal focusing lenses of the Gemini beam splitter. Furthermore, reference images of a fiduciary slide were used in postprocessing to correct translational misalignments at the pixel level (Supplementary Fig. 14b). Widefield imaging was performed on an inverted microscope (Zeiss Axio Observer) at 50x magnification (0.55 NA EC Epiplan NEOFLUAR objective) under an excitation power density of 13 kW/cm$^2$ (MDL-N-980-8W). The camera was controlled with the manufacturer's software, HC Image Live. Single-channel UCL images were recorded at 66 fps, and the full pixel array (2,048x1,024, no binning) was used. For correlated electrical and dual-channel optical video recordings, the camera was operated in external start trigger mode at 50 fps using a 512x128 pixel subset per channel and 2x binning. In this mode, the camera was set to begin acquisition after receiving a 3.3 V trigger pulse from an Arduino Leonardo. This Arduino also initiated the EPG acquisition via a near-synchronous mouse click in the NemAcquire software. This method results in an average lag of 64 ms (±18 ms) between the onset of acquisition in the two data streams. This corresponds to an average uncertainty in the alignment of the data streams equivalent to a single image frame (Supplementary Fig. 13). The maximum offset was two frames in either direction of the mean. Between one and five videos were acquired for each worm before the animal was ejected from the channel. Images of the empty channel collected under the same exposure conditions were used for background subtraction (Supplementary Fig. 14a). Average pixel intensities in the terminal bulb regions of the red and green channels were extracted with built-in morphological image processing functions in MATLAB (Supplementary Fig. 14c-e). A DC offset was applied to the voltage trace in each event (Supplementary Fig. 15). See Supplementary Fig. 16 and the Supplementary Information Section 7 for exclusion criteria applied to the optical and electrical datasets.

**Data availability**

Source data for Figs. 1-4 are available via the Stanford Digital Repositories (SDR) service at https://doi.org/10.25740/ff923hb3417. All other source data are available upon reasonable request.

**Code availability**

The code used to extract ratiometric data from fluorescence spectra and segment images rely on prebuilt MATLAB functions and is available upon request.

**Acknowledgements**

J.R.C. and A.S. acknowledge financial support from the National Science Foundation (NSF) through the Graduate Research Fellowships Program. J.A.D is a Chan Zuckerberg Biohub – San Francisco Investigator and acknowledges funding from the Chan Zuckerberg Biohub San Francisco, as well as from the National Institutes of Health (NIH) under grant No. 1DPAI15207201, and from the NSF under grant No. 1933624. M.B.G acknowledges funding from the NIH under grant No. R35NS105092. C.A.M. acknowledges funding from the Wu Tsai Neurosciences Institute at Stanford University. C. Siefe received support from an Eastman Kodak fellowship. C. Shi acknowledges funding from the U.S. Navy NDSEG Fellowship under BAA #N00014-22-S-B001. X.W.G. and A.P acknowledge financial support from the Army Research Office under grant no. W911NF2020171. A.P.'s time for reviewing and editing the written manuscript was supported under the auspices of the U.S. Department of Energy by the Lawrence Livermore National Laboratory (LLNL) under Contract DE-AC52-07NA27344 (LLNL review and release number LLNL-JRNL-854745-DRAFT). TEM/SEM imaging was performed at the Stanford Nano Shared Facilities (SNSF), supported by the National Science Foundation under Award ECCS-2026822. This work would not have been possible without technical support from Z. Liao. We also thank B. Myers and A. Jayich for providing the confocal MATLAB code, H. Ji for consultations on data analysis, B. Ogulande for acquiring the Raman spectrum, M. Cano for counting worm progeny and R. Chen for suggesting the use of DMAP.

**Competing interests**

The authors declare no competing interests.

**Author contributions**

# Extended Data Figures

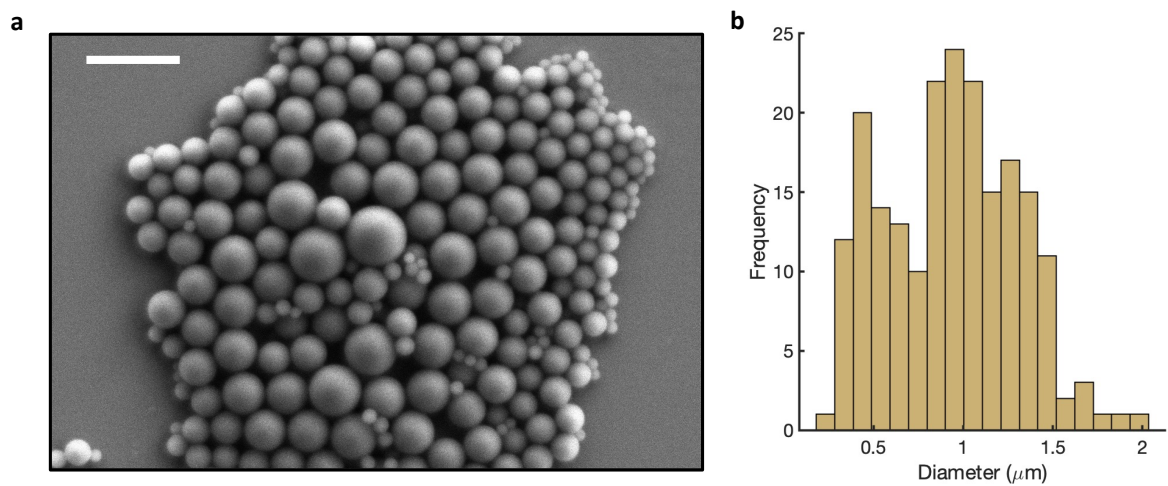

**Extended Data Fig. 1** Microgauge size measured via SEM

**a)** A SEM micrograph of a monolayer of uncoated microgauges on silicon. Scale bar, 3 μm. **b)** The distribution of N = 202 microgauge diameters (mean ± s.d. = 935 ± 367 nm), repeated for clarity from Fig. 1b).

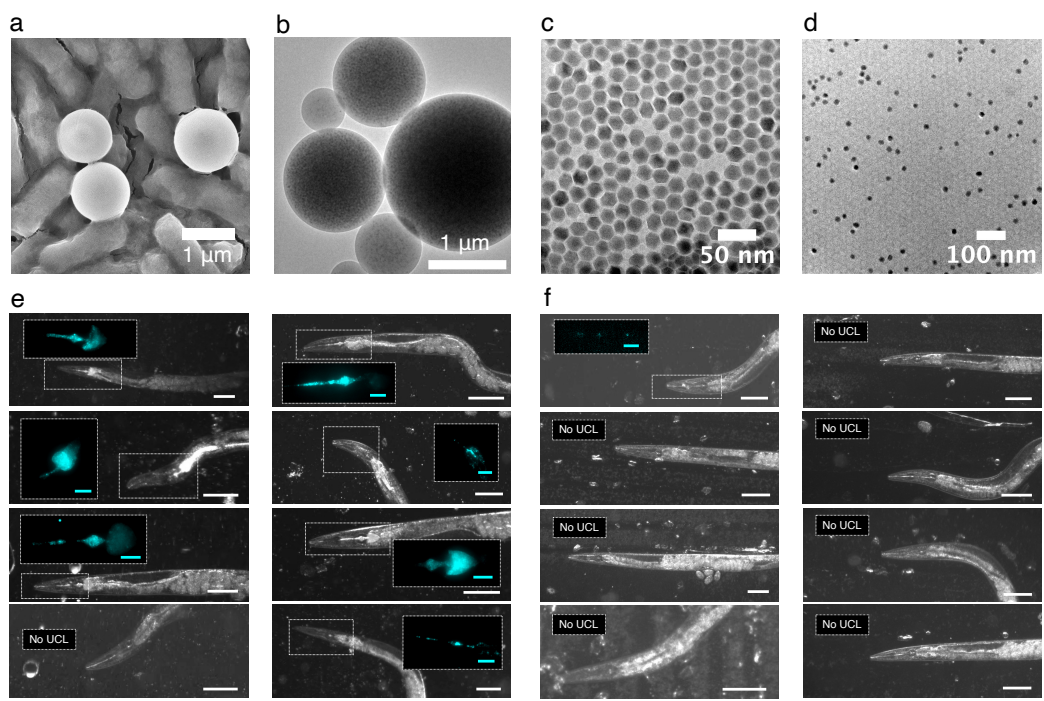

**Extended Data Fig. 2** The effect of sensor size on pharyngeal accumulation efficiency

**a)** An SEM micrograph of microgauges on *E. coli* for size comparison. **b)** A TEM micrograph of a microgauge. TEM micrographs of core@shell $NaY_{0.8}Yb_{0.18}Er_{0.02}F_4@NaYF_4$ UCNPs (**c**) before and (**d**) after wrapping with PMAO. Wrapped nanoparticles were dropcast from water. A series of bright field images of levamisole-treated worms after one hour incubation with (**e**) microgauges or (**f**) singly dispersed PMAO-wrapped UCNPs. Scale bars are each 100 μm. False-colored UCL images

(cyan) of the boxed pharyngeal regions are displayed in the inset of each picture, when applicable. Inset scale bars are each 20 µm each. The anterior of each animal is to the left.

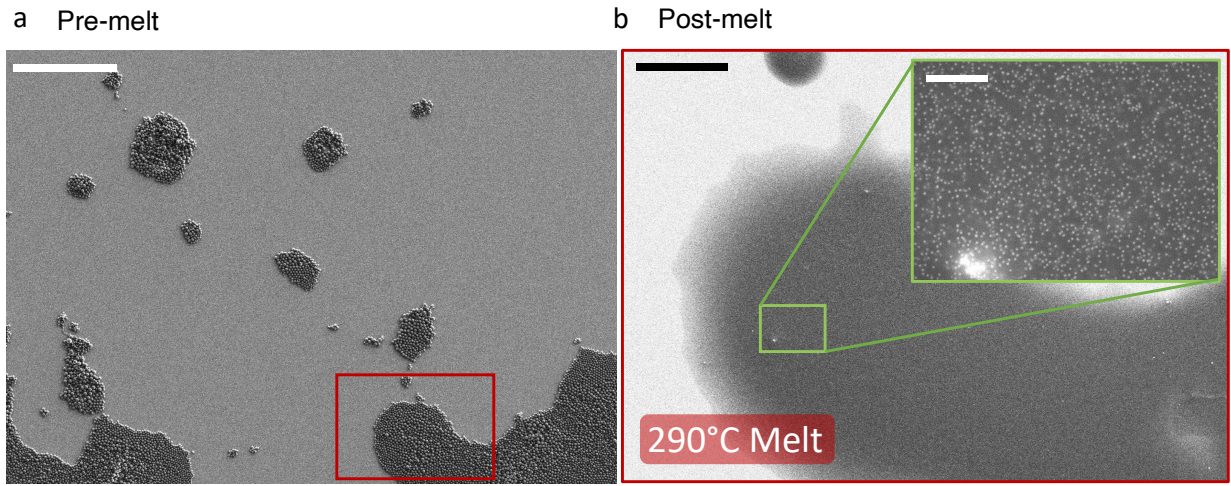

**Extended Data Fig. 3** Microgauge monolayers before and after melting

**a)** An SEM micrograph of microgauge monolayers on hydrophilized silicon. Scale bar, 50 µm. **b)** The same region after melting (30 minutes), with a higher resolution inset to show nanoparticles distributed evenly across the surface. Scale bars, 10 µm, 500 nm (inset). Clusters like the one shown in the bottom left of the inset were easily visible from confocal UCL maps and were avoided during the calibration.

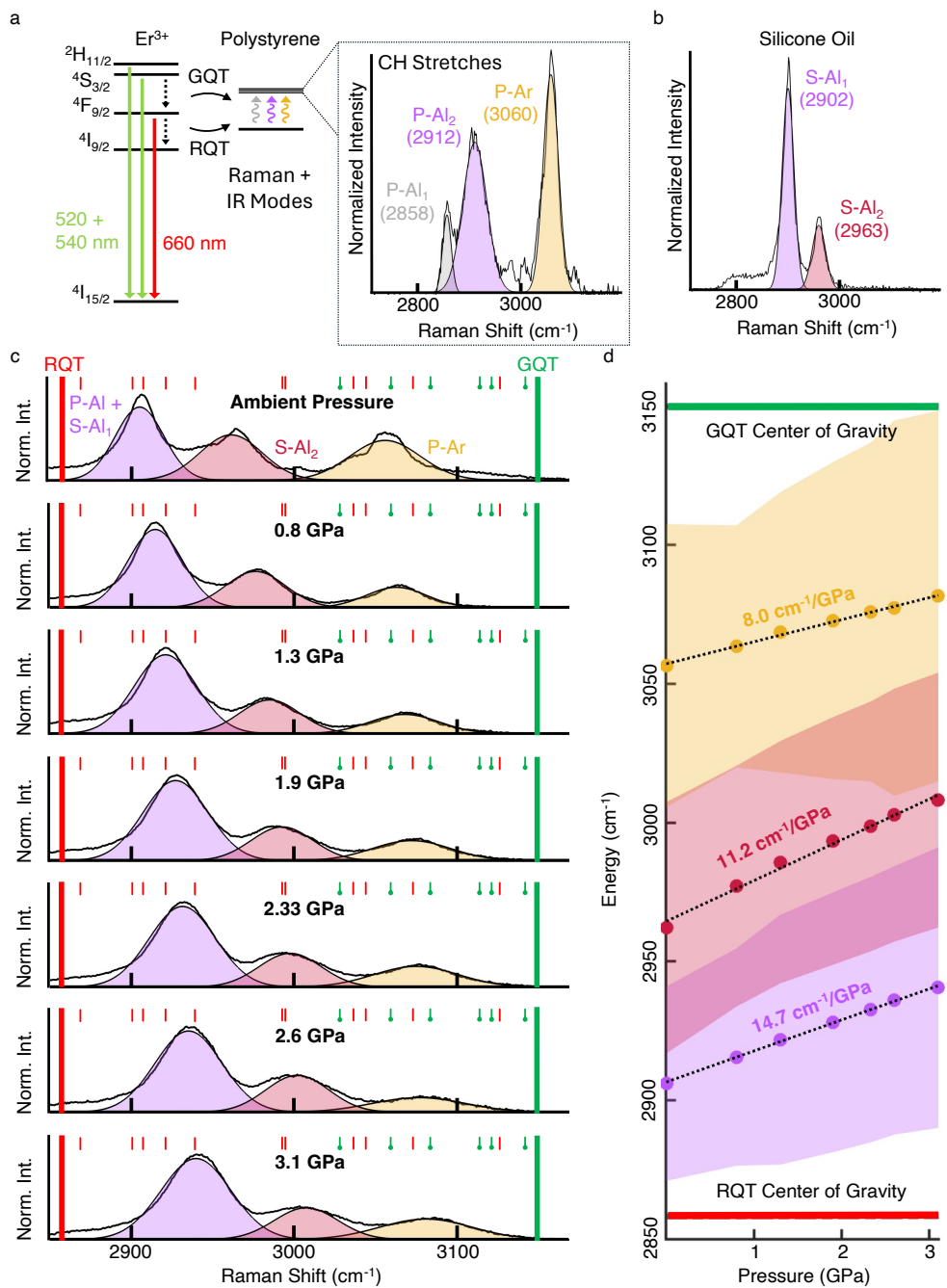

**Extended Data Fig. 4** Er-Polystyrene energy overlap

**a)** A simplified energy diagram of $Er^{3+}$ that summarizes matrix quenching. Solid lines indicate radiative transitions, and dotted lines indicate non-radiative Energy Transfer (ET) to the aliphatic (gray and purple) and aromatic (yellow) C-H stretching modes on polystyrene. The corresponding Raman spectrum for polystyrene microbeads without nanoparticles at room temperature and ambient pressure (right). **b)** Raman spectrum for silicone oil (room temperature, ambient pressure). **c)** Raman spectra of polystyrene microbeads and silicone oil between ambient pressure (top) and 3.1 GPa (bottom). Fitted curves are filled in using the same color scheme as in panels (**a**) and (**b**). P-Al$_1$, P-Al$_2$ and S-Al$_1$, too close to deconvolve at all pressures, are fitted to a single peak. Thick vertical lines indicate the centers of mass of the red quenching $^4F_{9/2} \rightarrow ^4I_{9/2}$ (2858 cm$^{-1}$) and green quenching $^4S_{3/2} \rightarrow ^4F_{9/2}$ transitions (3150 cm$^{-1}$), respectively (values obtained for LaF$_3$ from Carnall et al.[60]). Smaller red and green (circular end cap) colored lines above the spectra indicate the transition energies between individual stark levels that are above and below the centers of mass, respectively. Values in cm$^{-1}$ are as follows; RQT: 2871, 2904, 2912, 2924, 2941, 2992, 2994, 3038, 3046, 3075, 3128; GQT: 3030, 3059, 3083, 3112, 3112, 3120, 3141 (values obtained for LaF$_3$ from Carnall et al.[60]). **d)** Peak energy (dots) and 2x standard deviations (colored area) of the fitted Gaussian. Dotted black lines are the respective linear fits of the peak energies with pressure.

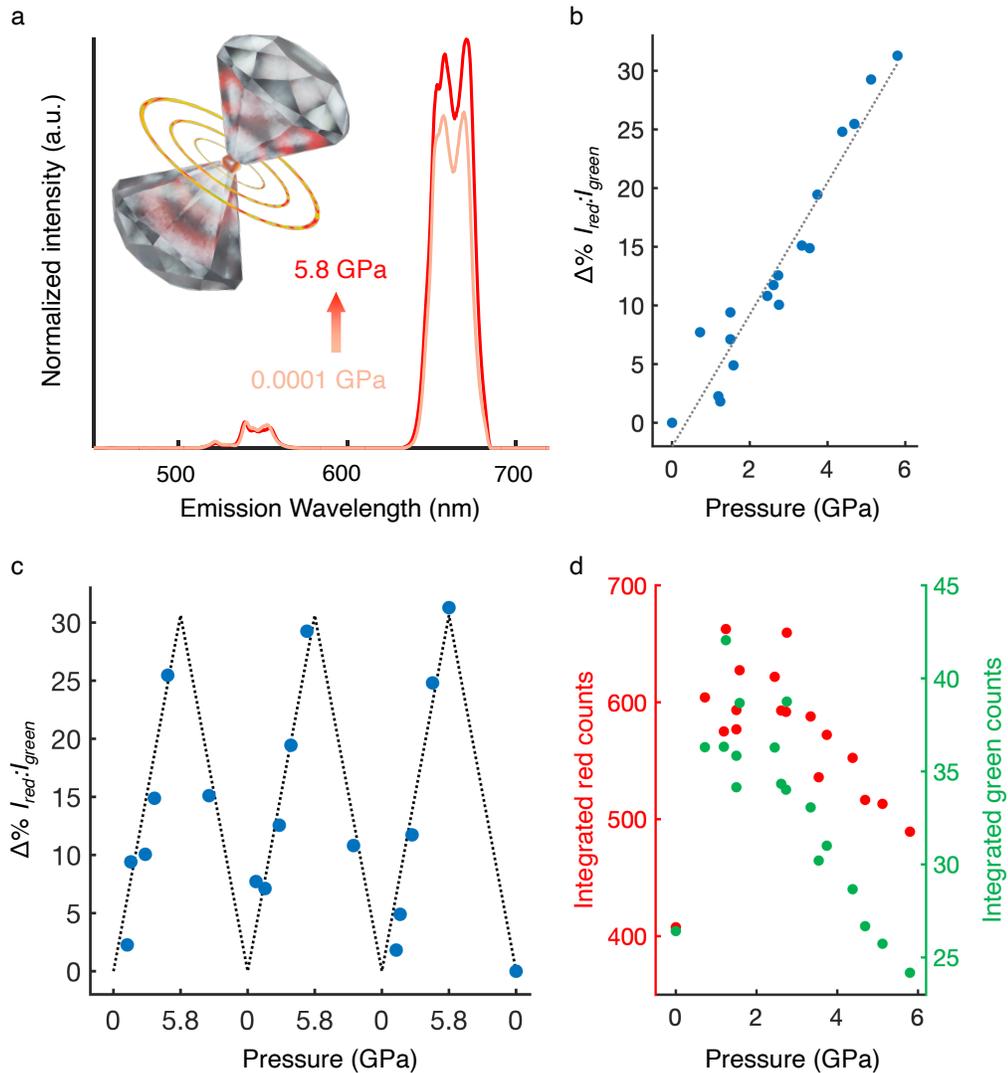

**Extended Data Fig. 5** I$_{Red}$:I$_{Green}$ increases with (quasi)hydrostatic pressure

**a)** Two spectra taken at either end of the pressure ramp showing the increase in relative red emission at high pressure. Both spectra are normalized to their respective green emission peaks. Inset, a schematic of the setup, wherein microgauges are compressed between two diamond culets. **b)** Ratiometric emission changes across three consecutive force loading-unloading cycles between ambient (0.0001 GPa) and high (5.8 GPa) pressure. The black line is the best fit line calculated in **c)**, where all ratiometric changes are plotted versus pressure regardless of loading or unloading status. **d)** The corresponding background corrected red (left axis, red) and green (right axis, green) UCL intensities as a function of pressure.

# Supplementary Information: Upconverting microgauges reveal intraluminal force dynamics *in vivo*

Jason R. Casar[7*], Claire A. McLellan[7], Cindy Shi[7], Ariel Stiber[7], Alice Lay[1], Chris Siefe[7], Abhinav Parakh[2,7], Malaya Gaerlan[3,7], Wendy Gu[4], Miriam B. Goodman[5*], Jennifer A. Dionne[6,7,8*]

[1]Department of Applied Physics, Stanford University, Stanford CA, 94305, USA

[2]Materials Engineering Division, Lawrence Livermore National Laboratory, Livermore CA, 94550, USA

[3]Department of Biology, Stanford University, Stanford CA, 94305, USA

[4]Department of Mechanical Engineering, Stanford University, Stanford CA, 94305, USA

[5]Department of Molecular and Cellular Physiology, Stanford University, Stanford CA, 94305, USA

[6]Department of Radiology, Stanford University, Stanford CA, 94305, USA

[7]Department of Materials Science and Engineering, Stanford University, Stanford CA, 94305, USA

[8]Chan Zuckerberg Biohub, San Francisco, San Francisco, CA 94158

*Address correspondence to: jcasar@stanford.edu; mbgoodman@stanford.edu; jdionne@stanford.edu

# Contents

# Section 1: *C. elegans* toxicity

## Laser heating assay

We adopted established bioassays for heat stress in *C. elegans* to evaluate the possibility that NIR illumination itself might be toxic to living animals. This assay relies on transgenic animals (TJ375 *gpIs1*[hsp-16.2p::GFP]) expressing GFP under the control of the promoter of the *hsp-16* heat-shock protein gene. This gene expression reporter generates visible fluorescence in the muscles of the pharynx 15-18 hours after heating to 35°C or above for 1-2 hours.[1] (*C. elegans* are healthy at temperatures between 15 and 26°C.) Replicating this finding, GFP fluorescence was stronger 16 hours after 2 hours at 37°C than 2 hours at 20°C (data not shown). Having established the utility of this assay as a heat-stress reporter, we analyzed young (day 1) adult worms for GFP fluorescence 16 hours after one of three conditions: 1) Control temperature (20°C), 2 hours; 2) Noxious heat (37°C), 2 hours, and 3) NIR (980nm, 13 kW/cm$^2$) laser irradiation, 1 minute. The last condition was performed on worms immobilized in the microfluidics devices used for EPGs and mechanical imaging. Control and noxious temperatures were achieved by placing animals in temperature-controlled incubators. As shown in SI Fig. 1, GFP expression was strongly upregulated in animals treated at 37°C for 2 hours. Based upon quantification of GFP expression in the head, the response to NIR irradiation was an order of magnitude lower. Treatment at 20° for 2 hours produced no detectable GFP expression. Thus, NIR laser irradiation seems to produce a mild heat stress response.

## Progeny Counting

The total brood size of adult worms is a useful window into toxicity. The general approach is to place single animals in one well of a 4-well plate, transferring them daily to a new egg-laying well for 3 days, and then to count the number of progeny produced by each parent worm. Assay plates were imaged twice on a flat-bed scanner, and progeny were counted by hand by an experimenter unaware of the treatment condition (SI Fig. 2). As seen in the zoomed-in green box, the size, shape, and contrast of the worms make them clearly distinguishable from bacterial food and debris. The second scan, taken 5' after the first one, further helped to differentiate debris or scratches from mobile worms.

## Section 2: Thermogravimetric analysis

Using fluorescence as a metric to compare the pharyngeal accumulation of PMAO-wrapped UCNPs (vs. microgauges) requires that we deliver an equal nanoparticle mass to animals in both conditions. We determined the mass fraction of $NaY_{0.8}Yb_{0.18}Er_{0.02}F_4@NaYF_4$ UCNPs in microgauges via thermogravimetric analysis (TGA, TA Instrument Q500). Samples were prepared as follows: 5.04 mg of sample was transferred to a flame-dried pan, heated at a rate of 20°C/min to 110°C, and held at that temperature 20 minutes to remove water. Sample mass was recorded as the pan was heated from 110°C to 550°C at a rate of 10°C/min under a nitrogen flow of 100mL/min (SI Fig. 3). To estimate how much of the remaining mass was due to residual, unoxidized carbon, we also tested polystyrene microspheres. This control was synthesized using an identical procedure (without UCNPs) and ramped under the same test conditions. The difference in residual mass was calculated to be 5.7%.

## Section 3: Confocal microscopy

### Confocal optical train

A schematic of the entire optical train is provided in SI Fig. 4. All components are mounted in a vibration isolation enclosure except those depicted in the inset, which are mounted on an external floating optical table. Except for the microscope's internal tube lens, all lenses are achromatic doublets with the appropriate anti-reflective coating (Thorlabs). Any unlabeled 45° lines represent silvered mirrors. We operated the 980 nm laser (Coherent OBIS) at 100 mW continuously. We measure a beam diameter of 0.7 mm exiting the laser and 1.75 mm entering the collimator (d) downstream of the lens pair. The collimator focuses the excitation onto a single mode fiber having a mode field diameter (MFD) of 5.3 µm (Thorlabs, P1-980A-FC-5) in order to clean the beam profile. Inside the enclosure, the beam profile is shaped by another lens pair. From this point, it passes through a 925-nm dichroic filter (FF01-Di01) to a Newport fast steering mirror (FSM-300-02). The FSM rasters the excitation beam across the sample (and also descans the UCL emission onto a common axis in the collection path). The addition of another lens pair allows us to expand the beam further to fill the back focal plane of the objective, and thus enhance spatial resolution. The objective is a cover glass (0.17 mm) corrected, 40x 0.95 NA Plan-Apochromat Zeiss objective with a back aperture diameter of 7.6 mm. We measure an overfill of 10 mm. The lateral resolution measured from the two channels is consistent (1.02 µm for transmission and 1.03 µm for reflection). The collection path deviates from the excitation path at the dichroic filter. We then filter all residual excitation light from the path with two shortpass filters (m,n). The collection path culminates in a 50:50 beamsplitter cube (Thorlabs, CCM1-BS013/M) that partitions the beam into the imaging and spectroscopy channels. Here, two identical single-mode fiber-collimator pairs function as fixed pinholes. Small thermal fluctuations of the stage can result in tens to hundreds of nanometers of drift, which can affect the power delivered to the sample (axial drift) or the force experienced by the excited region of the film (lateral drift). We minimize thermal drift by holding the AFM head at 26°C (± 0.02°C); it takes approximately 12 hours for the temperature controller to achieve this equilibrium inside the sealed vibration isolation chamber.

### Co-alignment for simultaneous imaging and spectroscopy

To guarantee that thermal drift did not affect the real-time coincidence between the confocal excitation spot and the center of the AFM tip-sample contact, we built two separate collection channels into the confocal optical train. The first was coupled to a spectrometer for mechano-optical calibration, as described previously. The second was coupled to a single photon counting device (SPCD) to monitor UCL counts during a spectral acquisition, and generate images between loading-unloading cycles. Near-ideal channel co-alignment was confirmed by overlaying diffraction-limited images of single nanoparticles. The average X and Y emission profiles of ten images taken over the course of 10 minutes on each channel were fit to a 2D Gaussian (SI Fig. 5). Finally, we determined the initial coincidence between the excitation spot and contact area for each film location by indenting the tip to our reference force point (0.4 µN) and observing the change in emission profile of the film. The center of the excitation spot is positioned at the center of the spherical contact area (SI Fig. 6).

## Section 4: Nanoindentation

### Compressive modulus characterization

We generated force-indentation curves from 50 microgauges according to methods described in the main text. We then fit each approach curve to a Johnson-Kendall-Roberts (JKR) model using a gradient descent method. A JKR model was chosen to account for the adhesive forces that produced a distinct "jump-to-contact" in the approach curves. From this analysis, we calculated an average modulus of 740 MPa with a standard deviation of 410 MPa (SI Fig. 7), which is within the range reported for microscale polystyrene in similar AFM indentation studies.[2–4] We probed microgauges between 0.18 and 2 µm in diameter, and did not observe any significant size (volume) dependence in the measured moduli (SI Fig. 7c). For small indentations, the parabolic tip used can be approximated as a sphere with radius $R_{tip} = 30\ nm$. We extracted radii for the microgauges from their accompanying height maps, $R_{mg} =$

$AFM\ Height/2$. Finally, we assumed that the Poisson's ratio of the microgauges was similar to that of polystyrene, $v_{mg} = 0.265$.

The assumed analytical form for indentation, $\mathbf{\delta}$, according to the JKR model[5] is:

$$\hat{\delta} = \frac{F_J^{\frac{2}{3}} - \frac{4}{3}(F_J F_{ad})^{\frac{1}{3}}}{(R_{eff} K^2)^{\frac{1}{3}}} \qquad \text{SI Eq.5}$$

where,

$$R_{eff} = \frac{1}{\frac{1}{R_{mg}} + \frac{1}{R_{tip}}}, \qquad \text{SI Eq. 6}$$

$$F_J = (\sqrt{F_{ad}} + \sqrt{F_{ad} + F})^2, \quad \text{and} \qquad \text{SI Eq. 7}$$

$$K = \frac{\frac{4}{3} E_{mg}}{1 - v_{mg}^2} \qquad \text{SI Eq. 8}$$

We estimated the adhesive force, $F_{ad}$, from the absolute value of the minimum of the baseline corrected approach curve following jump-to-contact. The exact onset of indentation, $\delta_0$, is not easily distinguishable in force-indentation curves of adhesive, deformable samples such as the ones tested here. However, within the JKR model, there is an analytical dependence between $\delta_0$ and an adhesive force such that $\delta(F) = \delta_0$ when $F_J(F) = 2.37 F_{ad}$. Although K could more accurately be written as, $K_{eff} = \frac{1}{\frac{1-v_{mg}^2}{\frac{4}{3}E_{mg}} + \frac{1-v_{tip}^2}{\frac{4}{3}E_{tip}}}$, K is a good approximation when $E_{tip} \gg E_{mg}$. This is the case for our silicon tip ($E_{tip}$= 200 GPa). A sum of squared residuals loss function, $\sum(\hat{\delta} - \delta)^2$, was used to converge on an estimate of the fit parameter, $K$. From trial and error, we found that with a learning rate of $5 * 10^{13}$ we could achieve convergence to $10^{-5}$ GPa for all samples in under 10,000 iterations. $E_{mg}$ was then extracted from the K value corresponding to the optimal fit according to SI Eq. 8. SI Fig. 7 shows a sample fit performed with this method and all 50 of the $E_{mg}$ values obtained from this fitting procedure. The linear least squares fit of compressive modulus versus volume has an unconvincing $R^2$ of 0.44 (SI Fig. 7c). Thus, we do not expect size to be significantly correlated to mechano-optical responsiveness.

## Contact pressure estimation

Having an estimate of the compressive modulus of the microgauge material allows us to estimate the maximal Hertzian contact stresses experienced by the microgauge film under various loading conditions, and compare it to the compressive yield stress of microscale polystyrene.[6] For a spherical indenter of radius, $R_{tip}$, the <u>maximum</u> stress experienced by the sample for a given uniaxial compressive load, $P$, is given by[7]

$$\sigma_{max} = \frac{\sqrt[3]{6PE_{eff}^2}}{\sqrt[3]{\pi^3 R_{eff}^2}} \qquad \text{SI Eq. 9}$$

And the mean contact stress is ⅔ this value.[7] In this case, we are assuming that the microgauge film being indented is flat with an infinite radius, and thus $R_{eff} \approx R_{tip}$, and that both the silica and silicon indenters are significantly stiffer than the film, $\frac{1}{E_{eff}} \approx \frac{1-v_{mg}^2}{E_{mg}} \approx \frac{1}{800\ MPa}$. Once again, we make the assumption that the indentation depth of the rounded parabolic indenter is small relative to its radius of curvature, so we can approximate it as a spherical indenter. Employing SI Eq. 9, we estimate that the maximal contact stress used in the mechano-optical characterization with the spherical silica tip is $\sigma_{max}$= 42 MPa. We employ this same analytical form to estimate the maximum contact stress experienced by the 1) microgauges and 2) *E. coli* within the grinder. To our knowledge, there is no experimentally derived stiffness for the grinder cuticle, so we estimate that it is similar to that of chitinous structures within animals that have been previously characterized with nanoindentation. These studies report average fitted moduli in the range of just under one GPa to over ten GPa.[8–11] We recognize that many factors affecting stiffness (such as the relative fraction of chitin, the nano-to-micrometer scale structural organization of chitin alone and in conjunction with other components, and the structure's porosity) are likely to differ between structures. However, because this range reflects structures having a variety of functions derived from an array of different species, we believe it is likely to be representative of the grinder cuticle as well. Based on the assumption that the cuticle stiffness is ~1 GPa and has a Poisson's ratio similar to that of polystyrene, the effective modulus is $E_{eff} \approx 410\ MPa$. Depending what part of the lumen is in contact during compression, the contact area is limited by either the average microgauge diameter or the width of the grinder's individual teeth (~200-500 nm, 375 nm average).[12] These assumptions yield average Hertzian contact stresses for 15.7 µN of 84 MPa and 155 MPa, respectively.

As mentioned in the main text, high pressure homogenizers operating under different conditions between about 50 and 200 MPa[13,14] can achieve the same thousand-fold *E. coli* inactivation rate measured by Vega and Gore (2017) *in vivo*.[16] The pressure-to-force conversion we employ to estimate a corresponding force range relies on estimates of the geometry and stiffness of the cuticle (discussed above) as well as *E. coli*, which fall in the range of tens of MPa to a few hundred MPa.[15] For simplicity, we assume that the bacteria is oriented with its long dimension normal to the grinder cuticle, and can thus be modeled as a 500 nm diameter sphere with a compressive modulus of 50 MPa. An average hertzian contact stress of 50 (or 250) MPa would require a compressive force of 4.2 (or 270) µN.

## Plastic deformation tests

We sought to confirm that the mechano-optical calibrations were performed within the elastic regime of the microgauge film. To do so, we replicated the same three-cycle loading conditions on a microgauge film sample and used scanning electron microscopy to visually compare the surface integrity before and after the indentation. A 30-second dwell time was used for each force actuation, and the sample was imaged without any conductive deposition layer to prevent interference with the compressive testing. No scarring or divotting was observed in the sample (SI Fig. 8). To confirm that it was possible to observe scarring with stresses exceeding the predicted yield strength, we applied three different forces, each at a different film location, with a sharper 30 nm radius silicon tip (SI Fig. 9). Although it exceeded the estimated yield stress of 100 MPa by a large margin, the first force (0.5 nN, $\sigma_{max}$= 400 MPa) did not produce plastic deformation. The latter two (1 nN, $\sigma_{max}$= 510 MPa; 2.5 nN, $\sigma_{max}$= 700 MPa) did.

## Section 5: Microgauge pressure sensitivity mechanisms

The trend towards redder $Er^{3+}$ UCL color at high pressure is consistently observed in co-Yb-Er-doped nanoparticles.[17–19] This raises the questions: What material changes are taking place at high pressure, and what transitions within the UC network are impacted by these changes? The energy transfer (ET) network[20] that results in red and green emission from near-infrared excitation is complex, with a number of photon- and phonon-mediated processes. It is logical to expect that pressure would impact donor-acceptor pair separations, lattice or matrix phonon energies, index of refraction and local symmetry. Each of these could in turn impact any number of steps along the red and green UC pathways, including multiphonon relaxation (MPR), energy transfer upconversion (ETU), cross relaxation (CR), energy migration, and radiative emission. ED Fig. 4d, demonstrates that the rate of green UCL loss with pressure (33%) is faster than the rate of red UCL loss (20%) between 2.75 and 5.8 GPa, suggesting that differential loss mechanisms could contribute to observed color changes. This discussion will highlight two mechanisms that, if consequential, would produce a qualitatively similar trend, $Er^{3+}$-ion cross relaxation and matrix quenching.

### Er-ion cross relaxation

Lattice strain will decrease donor-acceptor pair separation, a critical parameter that is known to enhance ET transitions that populate the red and green emitting levels but also increase the cross relaxation and energy-migration-to-surface processes that depopulate them.[20–22] Several studies suggest that cross relaxation is an efficient mechanism for quenching green emission in Er-doped UCNPs. For example, Rabouw et al. showed that increasing erbium concentration from 0.1% to 2% enhances green quenching rates in β-NaYF$_4$ by three-fold, but has a relatively minor effect on red quenching rates.[23] The green quenching ET transitions Er ($^2H_{11/2} \rightarrow {}^4I_{9/2}$): Er ($^4I_{15/2} \rightarrow {}^4I_{13/2}$) and Er ($^2H_{11/2} \rightarrow {}^4I_{13/2}$): Er ($^4I_{15/2} \rightarrow {}^4I_{9/2}$) have an energy mismatch of 245 and 215 cm$^{-1}$, respectively. Compare this to the 2248 cm$^{-1}$ mismatch of the Er ($^4F_{9/2} \rightarrow {}^4I_{13/2}$): Er ($^4I_{15/2} \rightarrow {}^4I_{13/2}$) transition, which is the smallest Er-Er cross relaxation process that quenches red emission to an acceptor in the ground state.[24] In SI Fig 12 we compare the expected nearest Er$^{3+}$ acceptor (NEA) separation, $\hat{R}_{Er_i-NEA}$, for each potential Er$^{3+}$ donor site in a cubic 10.9 nm diameter NaYF$_4$ core to the critical Förster radius for Er$^{3+}$-Er$^{3+}$ cross relaxation calculated by Rabouw et al. (R$_0$ = 0.92 nm). In this analysis we assume an independent and homogenous probability, P = 0.01, that FCC acceptor sites are occupied by Er$^{3+}$, the lattice parameter is 5.51 Å,[17] the compressive modulus of the linear, isotropically elastic lattice is E = 272 GPa,[18] and the effects of unequal bond lengths are negligible. When the lattice is compressed hydrostatically from atmospheric pressure to 5.8 GPa, the proportion of donor sites with $\hat{R}_{Er_i-NEA}$ less than R$_0$ increases from 52% to 64% (SI Fig. 12 inset). This implies that, on average, more donor sites excited to the green emitting state would experience a cross relaxation rate that "outpaces" radiative emission. Er$^{3+}$-Yb$^{3+}$ cross relaxation is another efficient green quenching mechanism in cubic Y$_2$O$_3$.[25] Dong et al. suggest that proximity induced enhancement of a Er$^{3+}$-Yb$^{3+}$ cross relaxation-mediated green-to-red population pathway in non-stoichiometric, synthetically contracted α-NaYF$_4$ contributed to a fifty-fold I$_{Red}$:I$_{Green}$ enhancement.[26] Furthermore, if we assume the ET rate from an excited Erbium donor has a FRET-type separation dependence, $W = \frac{C}{R^6}$, then the maximum pressure employed in our UCL DAC experiment (5.8 GPa) enhances the ET rate by 13.8%. These data suggest that cross relaxation is a mechanism which disproportionately depopulates the green emitting manifold of Er$^{3+}$ donors, and which can be enhanced in a compacted lattice due to global ET rate enhancements.

## Polystyrene quenching

In addition to cross relaxation, the green-emitting $^4S_{3/2}+^2H_{11/2}$ and red-emitting $^4F_{9/2}$ states can be depopulated via ET to high energy vibrational modes immediately outside the nanoparticle. These processes are operative across larger donor-acceptor separations because of the relatively large oscillator strengths of the allowed vibrational transitions of the surrounding matrix.[23] Rabouw et al. found that the $^4F_{9/2}$ decay rate was significantly reduced in predominantly aromatic (versus aliphatic) solvents, but the $^4S_{3/2}$ decay rate of nanoparticles with identical erbium concentrations was robust to solvent character. In a similar vein, Fischer et al. determined that the ratio of the surface quenching rate to the intrinsic depopulation rate (ie, in nanoparticles with sufficiently thick shells) is higher for the red emitting state (1.1) than it is for the green emitting state (0.8).[27] These results suggest that the solvent - or the chemical environment more generally - differentially quenches red emission at atmospheric pressure. In support of these findings, we show in ED Fig. 5 that the Raman-active aliphatic -CH stretch of polystyrene (~2912 cm$^{-1}$) has good overlap with the center of mass of the broad red quenching $^4F_{9/2} \rightarrow ^4I_{9/2}$ transition (labeled RQT, ~2850 cm$^{-1}$) tabulated for LaF$_3$ by Carnall et al.[24] and measured directly for flourindate glass via MWIR fluorescence spectroscopy.[28] The aromatic -CH stretch of polystyrene (~3060 cm$^{-1}$) has slightly worse energy matching with the center of mass of the broad, green quenching $^4S_{3/2} \rightarrow ^4F_{9/2}$ transition (labeled GQT, ~3150 cm$^{-1}$), which again is tabulated for LaF$_3$ by Carnall et al.[24] and has been measured for fluoroaluminate-tellurite glass via MWIR fluorescence spectroscopy.[29] Although our vibrational spectra were acquired via Raman spectroscopy (see Methods subsection "Diamond anvil cell preparation for Raman and UCL measurement"), these transitions are also IR active.[30]

Although polystyrene's vibrational modes couple more strongly to the RQT at atmospheric pressure, they will broaden and blueshift at elevated pressure, improving resonance with the GQT and potentially enhancing $I_{Red}:I_{Green}$. From atmospheric pressure to 3.1 GPa , the polystyrene aromatic peak blueshifts by 25 cm$^{-1}$, towards the center of mass of the GQT (ED Fig. 5c-d). Pressure induced broadening may enhance resonance further. This transition is also a prominent $^4F_{9/2}$ inflow, so its potential to enhance $I_{Red}:I_{Green}$ at high pressures is twofold. Over this same range, the aliphatic peak(s) blueshift by 35 cm$^{-1}$, away from the center of mass of the RQT. The effect of this shift on RQT resonance is less clear because the RQT is likely to broaden at high pressures. Future pressure dependent MWIR fluorescence studies would clarify the trend in red emission quenching. It is worth noting that matrix quenching may have contributed to the $I_{Red}:I_{Green}$ enhancements measured in previous DAC experiments despite the absence of polystyrene, because the silicone oil pressure medium exhibits a qualitatively similar trend in its resonance with the two quenching transitions (ED Fig. 5c-d). Non-radiative outflow from the green emitting state could be stimulated further by the simultaneous densification of these vibrational modes. Using the Tait equation for a polystyrene glass at 20°C (SI Eq. 4), we estimate that this density should increase by 4.5% over only 200 MPa (the pressure range for which the equation of state is strictly valid):

$$\rho_P = \frac{\rho_0}{1 - C \cdot ln(1 + \frac{P}{B(t=20)})} \quad \text{SI Eq. 4}$$

where $C = 0.0894$, and $B(t = 20°C) = 3.267$ kilobar.[31,32] Overall, these data suggest that green emission may experience more efficient matrix quenching at high pressure.

Other mechanisms

There is also precedent in the literature to suggest that pressure mediates UCL response by distorting local symmetry, increasing the index of refraction (IOR) of the polystyrene matrix, and blueshifting lattice phonon modes. Wisser et al. show that hydrostatic pressures below ~2 GPa increase UCL intensity and decrease UCL lifetime in α-NaYF$_4$, but strictly decrease UCL intensity in β-NaYF$_4$, a distinction they explain in terms of local crystal field symmetry breaking.[33] We note that, although the actual point group for Ln$^{3+}$ centers in the cubic lattice is likely not octahedral at the doping concentrations we use, it is still likely to be centrosymmetric (D$_{4h}$) at atmospheric pressure.[34] UCL radiative emission rates are also positively dependent on the IOR of the suspension medium.[23] For polystyrene, IOR will scale linearly with hydrostatic pressure up to ~200 MPa, before dampening in response at higher pressures.[35,36] Finally, pressure will affect peak energies and widths for lattice phonon modes, but the impact this has on UCL is perhaps marginal. Runowski et al. measure a ~17 cm$^{-1}$ blueshift in the phonon spectrum of cubic-phase SrF$_2$:Er,Yb over 5.29 GPa, but conclude it is too small of a shift to explain the red and green lifetime reduction for Er$^{3+}$ UCL that they observe over that same pressure range.[37]

## Section 7: Electrical and optical measurement of pharyngeal function

### Signal synchronization: Trigger setup and lag

Two-channel (red, green) upconversion luminescence videos were acquired from ingested microgauges according to methods described in the main text in tandem with electrophysiological records of pharyngeal muscle activity (aka electropharyngeograms or EPGs). The data streams were synchronized using an Arduino Leonardo to trigger acquisition on both the camera (ORCA Flash4.0) and the EPG microfluidics chip. The camera was operated from a desktop PC (Dell Precision T3610, "camera computer") and the EPG and Arduino were operated from a laptop (Lenovo ThinkPad P1, "EPG computer"). The camera was operated in External Start Trigger mode, while the mouse cursor of the EPG computer was positioned to start the acquisition in NemAcquire, a software application provided by the EPG chip manufacturer (inVivo Biosystems). Once switched on, the Arduino delivered a TTL pulse via a BNC-to-SMA cable to initiate the camera acquisition and a left mouse click trigger (Mouse.click) via a USB cable. When operating the ORCA Flash4.0 in this triggering mode with an exposure time below 100 ms, the first exposure frame is dropped (Orca Flash4.0 Manual). This single dropped frame introduces a 20 ms lag between the receipt of the TTL trigger and the exposure of the first valid frame. There is a relatively negligible jitter of 30 μs. There is also a lag in the EPG computer's response to a mouse click event; because of the nature of the task scheduler in the Windows 10 OS, this lag is non-deterministic. Below we discuss how this jitter results in a maximum two-frame uncertainty in the true offset between the recording streams.

## Lag estimation

We used a simple method to measure the lag between the start of the camera and EPG acquisition (SI Fig. 13). To determine when a simultaneous electrical and optical signal would be registered on both recordings, we positioned an LED so that its emission would be captured by the camera and placed the wires driving the LED close enough to the chip's electrodes to generate an electrical signal. For each trial, we triggered a dual acquisition as above, and then manually switched the LED on and off at nearly regular intervals within the 60-second acquisition (50 Hz on the camera, 500 Hz on the EPG). In SI Fig. 13b, we plot the average pixel brightness and raw voltage time courses from one of the ten trials. In every trial, the relative time at which the camera registered the first optical signal preceded the relative time at which the EPG recording reflected the voltage change. Across 10 trials, we measured an average lag of 64 ms ±18 ms (min: 20 ms; max: 100 ms) (SI Fig. 13c). The temporal resolution (20 ms) of these measurements is limited by the camera frame rate (50 Hz). Thus, the average lag corresponds to three frames, with a one-frame uncertainty. We offset the optical time course by the average lag (three frames) when generating the event-triggered averages so that the maximum uncertainty in the coincidence with R would never be more than two frames in either direction. To confirm that a single lag offset could be applied to the entire time course without introducing systematic error within the trial, we also explored the correlation between the elapsed time and change in lag within individual trials (SI Fig. 13d). Any dropped frames (or voltage measurements) would result in an increase (or decrease) in the lag. We fit the change in lag relative to the first measured lag against the elapsed time within the trial according to a linear model with a fixed intercept of zero. Although there was a statistically significant correlation ($p = 5 * 10^{-5}$) between the two, the slope was only 0.17 ms/s (SE = 0.04 ms/s). This means that over the course of the 60 s, 3000 frame recording, lag between the optical and electrical signals is expected to increase by only 10.2 ms. The effect of this expected error on our measurements is negligible since it is roughly half of the acquisition time for a single imaging frame.

## Event-triggered averages

### Optical data processing

We converted raw UCL videos into time series of the total red and green emission from the terminal bulb of ingested microgauges in worms held in the EPG chip as follows. The general strategy was to locate microgauges in the red channel (which is the brighter of the two UCL channels) and to analyze the total intensity in both channels in this region of each frame. To reach this goal, in each frame, the red channel was binarized using a threshold pixel intensity calculated by the Sobel method and manually scaled to account for the variations in microgauge accumulation in the anterior intestines of individual worms. This binarized image was dilated with an octagonal kernel and eroded with a diamond kernel and kernel sizes were adjusted manually in order to achieve the best possible separation among regions of interest (ROIs) in the terminal bulb (imaging target), anterior intestine, and anterior pharynx. Images were filtered and bridged (to recover missed targets) using a 3D cuboidal kernel of size [x = 1, y = 1, frame = 5]. Of the three ROIs, the intestine was generally the largest and this size differential allowed the analysis to focus on the ROI associated with the terminal bulb. In a minority of frames, the intestinal ROI obscured the terminal bulb, leaving a gap in the time-series of UCL emission from the pharyngeal bulb. These were bridged, where possible, using the aforementioned 3D kernel. We measured the UCL intensity in the red and green channels for the terminal bulb ROI, generating a time-series of $I_{Red}:I_{Green}$ emission from microgauges transiting through the terminal bulb of the pharynx. We adjusted the resulting time series for the temporal lag between camera and EPG acquisition by culling three frames from the beginning of the optical time series, and a corresponding 60 ms worth of data from the end of the electrical time series.

### EPG data processing

To average electrical signals associated with a single pharyngeal pumping cycle, EPG traces were divided into individual pump events as follows. First, we identified a time window that would include the excitation or E phase, the relaxation or R phase, the interval in between these signals and an equal amount of time before the E phase and following the R phase. Based upon the average pump duration (104 ms rounded to 100 ms) and the average pumping frequency (0.88 Hz), the resulting time window was 460 ms in duration (SI Fig. 15). We refer to the span of time from 20 to 180 ms as the post-R epoch of the event window. Analogously, the span from -280 to -120 is labeled as the pre-E epoch. A 20 ms buffer was included to allow for the voltage of each transient to reach baseline. Next, voltages measured in the two epochs were averaged and used to scale the entire event window. For worms that were recorded inverted in the channel (ie, with their pharynx facing the back electrode), we then applied a scaling factor of -1 to the zeroed voltage trace. This baseline shift and sign correction were the only data processing steps performed on the electrical time series.

The quality of the voltage trace was used as a proxy for event quality that could be calculated independently of the optical time series. Not only did this help us identify and remove falsely flagged R events, but it also reflects how stably the worm pharynx was seated in the channel. Low-frequency signals with magnitudes comparable to that of R could indicate motion parallel to the immobilization channel or perpendicular to it, as well as background electrical interference. As a measure of event quality, we calculate the root mean square error of the baseline corrected voltage in two windows at either end of the event window. We call these 80 ms periods the RMS windows (SI Fig. 15). The positioning of the RMS window in the post-R epoch (100 to 180 ms) was chosen to allow for any R2 transients to reach a baseline voltage level. The RMS window in the pre-E epoch (-280 to 200 ms) was chosen so that it was equal in width to the other RMS window and could account for pump durations that were longer than the average. In other words, choosing to buffer the RMS window such a distance from the signal epoch prevents pump-inducing voltages from being counted in the RMS error in all recorded events. An RMS value greater than 20% of the magnitude of the R peak resulted in event exclusion (see Exclusion Criteria).

As a final note, we chose this simpler strategy to estimate event quality because it required fewer assumptions. Attempts to isolate the RMS of the "noise" from the pump-inducing signals via high pass filtering required us to make arbitrary assumptions as to frequency of the noise and resulted in values that still correlated with the magnitude of the signal in the pre-E and post-R epochs. Strategies based on polynomial fitting or moving averages suffered similar limitations. The relative magnitude of R provides an independent metric of signal that should be much larger than the magnitude of any feature in the pre-E or post-R epochs in which there is no pharyngeal muscle activation.

Event-triggered averaging of optical and electrical signals

To generate the event-triggered averages, we isolated individual pumping events from the EPG and the corresponding recording of $I_{Red}:I_{Green}$ in the terminal bulb of the pharynx, which is proportional to the force applied to microgauges in this location. The peak of the R phase was set as $t = 0$ and the time axis was normalized to the interval in between the E and R phases. The force recording was normalized to the $I_{Red}:I_{Green}$ value or "ambient force" observed during the E phase ($t = -1$), yielding $\Delta\%I_{Red}:I_{Green}$ across the event. Using an emission ratio links these measurements to force via our AFM-confocal calibration and circumvents the need to account for differences in collection efficiencies between widefield optics used to measure microgauge activity *in vivo* and the confocal optics used to calibrate mechanosensitivity, if present.

## Exclusion criteria

Several factors affect the quality of the imaging and EPG data streams obtained from individual animals loaded into the EPG chip. These include variations in microgauge ingestion, the need to rapidly load each worm into the EPG chip, and the quality of the immobilization. In total, we collected 92 one-minute, 3000-frame videos from 74 individual worms. However, many videos seemed to lack sufficient microgauge accumulation in the pharynx, as evidenced by low green emission. This situation likely reflects that ingested microgauges are continually transported from the pharynx to the intestines from the moment animals are removed from the feeding plate and loaded into the EPG chip. Delays in this process and in laser alignment result in few, if any microgauges being present in the pharynx when video acquisition is started (SI Fig. 16a). Additionally, some worms would slip within the immobilization channel, independent of the stability of the initial immobilization (SI Fig. 16d). Occasional anterior body contractions or lateral motions of the pharynx resulted in sudden, large movements of the terminal bulb that impaired image segmentation. Finally, there were several instances where the noise in the EPG signal exceeded the ability of the manufacturer's NemAnalysis software to locate the E/R transients. Forty-nine (49) videos that showed evidence of low signal, instability of the worm or pharynx in the EPG chip, or excessive noise in the EPG signal were excluded before any segmentation was attempted.

Of the remaining 43 videos, 22 were segmented using our image processing pipeline. Videos could be segmented if image analysis parameters could be chosen to capture the terminal bulb and exclude the anterior pharynx and intestines in all frames (SI Video 6). In the 21 videos that could not be segmented, the terminal bulb was too close to the intestinal signal (SI Fig. 16b), the isthmus had a significant signal for a large portion of the video, or the terminal bulb signal was too low compared to that of the intestines and the anterior pharynx. Two hours of parameter optimization were attempted on each video before attempts at segmentation were abandoned. This decision was independent of any knowledge as to the magnitude or direction of $I_{Red}$:$I_{Green}$ change in the terminal bulb, since segmentation is required to obtain that information.

The quality of the segmentation was rated by three volunteers. In brief, outlines of the segmented pixels were overlaid on each video and coded videos were viewed by volunteers who rated the videos on a scale of 1 to 5 based on three criteria: 1) the frequency and magnitude with which terminal bulb signal breached the segmentation outline during a pump (SI Fig. 16c), 2) the average proportion of terminal bulb signal that was excluded from the outline across all frames, and 3) the proportion of the total segmented pixel region that comprised either the intestines, isthmus or anterior pharynx. A score of one (1) represented a segmentation with frequent breaches that did not fully capture the terminal bulb and which often captured signal outside of the terminal bulb, while a score of five (5) represented a segmentation that fully and exclusively captured the terminal bulb signal. Individual scores were averaged, and all videos receiving a score of two (2) or less were excluded. These threshold criteria were not provided to scorers. Seven (7) videos were excluded based on this criteria, leaving fifteen (15) videos with high-quality segmentation. Imaging pipelines that make use of machine-learning based tools for image segmentation are emerging as a more robust tool than intensity-based segmentation and could harvest more information from this data set in the future and from additional similar data.

Although the selection of a 460 ms event window was appropriate given the pump durations and pumping frequencies exhibited by the vast majority of the worms, it posed an issue for individual worms pumping faster than 2 Hz on average. These individuals would exhibit R (and E) transients in nearly all of the pre-E (and post-R) epochs. These events would then be flagged for exclusion based on our previously determined criteria. Thus, we excluded four data sets that had an average pumping frequency above 2 Hz rather than including the negligibly few events where there was no signal interference from the previous or subsequent event. A second frequency exclusion criteria put the minimum acceptable pumping rate at 0.167 Hz, or 10 pumps in 60 seconds. It was assumed that any worm pumping slower than this in the presence of serotonin was not representative. Four (4) additional data sets were excluded based on this criteria, leaving the final remaining total of seven data sets. Each came from a separate worm, and spanned three of the five days of study. From the seven worms, a total of 61 events were excluded based on the RMS criteria, leaving a final event total of 185. As an example, all 12 of the 30 total pumping events corresponding to Fig. 4c.iv and SI Video 6 that did not pass RMS quality control are displayed in SI Fig. 16e.

# Section 8: Supplementary Figures

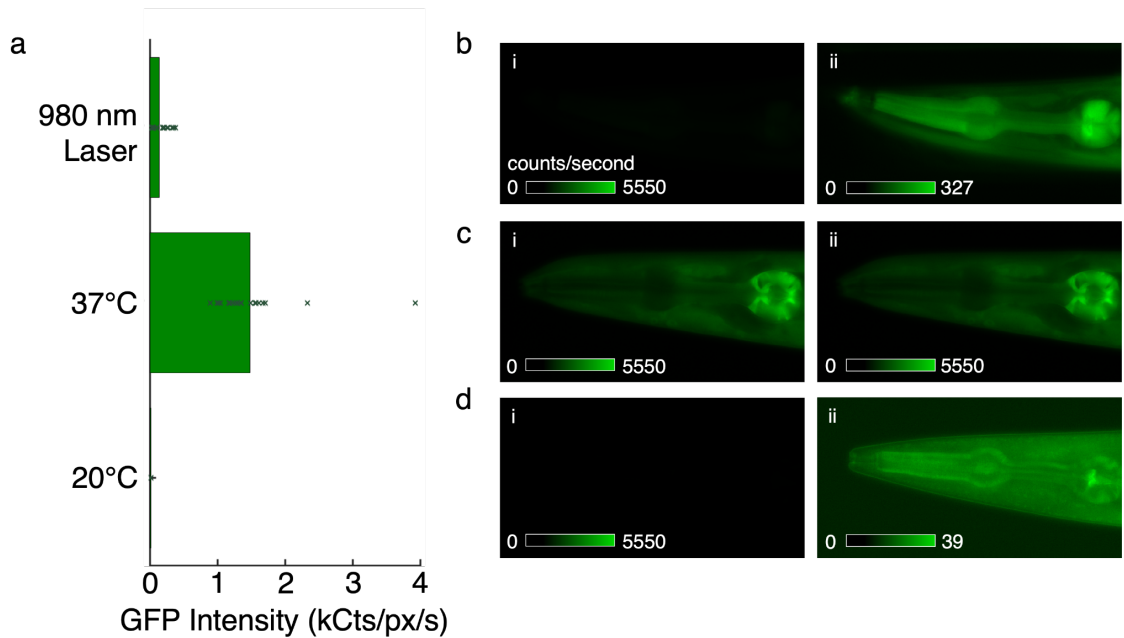

**SI Fig. 1: 980 nm laser irradiation induces modest thermal stress**

**a)** Average GFP intensities in the heads of TJ375 *hsp-16.2::GFP* transgenic worms 16 hours after a one-minute exposure to a 13 kW/cm$^2$ 980 nm laser (top, 143 Cts/px/s), a 2 hour incubation at 37°C (middle, 1483 Cts/px/s), a 2 hour incubation at 20°C (bottom, 17 Cts/px/s) and N=21 for all three conditions. **b-d)** Representative fluorescence images of GFP expression in the three conditions at the same scale (980 nm laser: **bi**, 37°C: **ci**, 20°C: **di**) and on separate scales (980 nm laser: **bii**, 37°C: **cii**, 20°C: **dii**). In the control condition, there is a significant luminal component from ingested bacteria. The contribution is small in the other two conditions, and the lumen appears relatively dark.

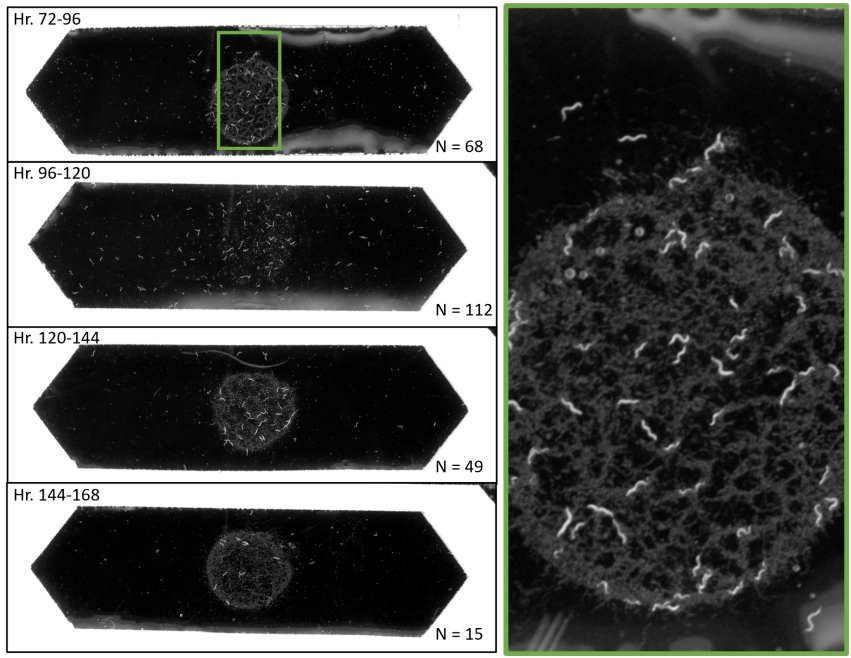

**SI Fig. 2: Egg-laying wells**

Scanned images of egg-laying wells that were used to count the number of progeny produced within each 24-hour period. All four images correspond to the same egg-laying individual, which belonged to the microgauge-fed test group. The hours since bleaching and the number of worms counted are labeled in the upper-left and lower-right corners of each image, respectively. The white boundary is a cut foam insert that discourages worms from crawling out of the arena.

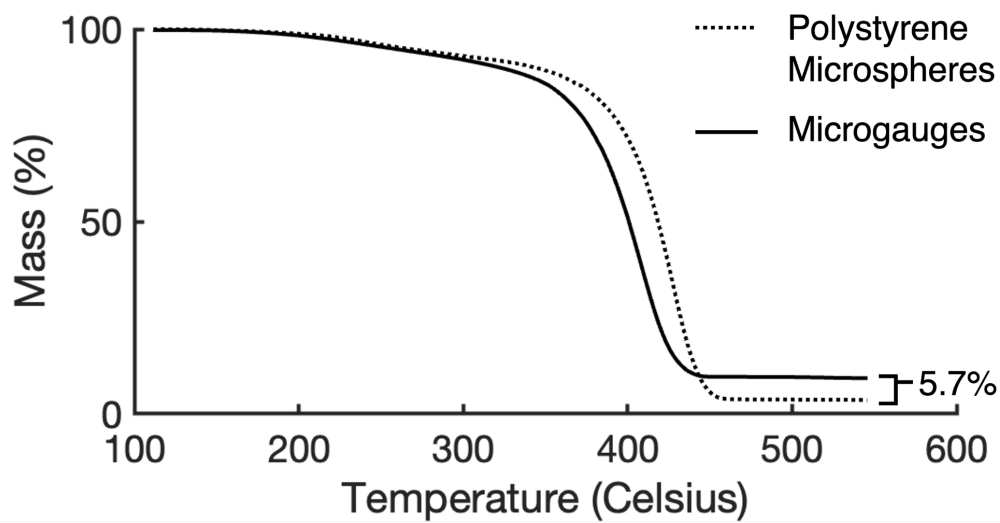

**SI Fig. 3: Thermogravimetric Analysis of UCNP mass fraction in microgauges**

A plot of residual mass as a function of temperature for polystyrene embedded $NaY_{0.8}Yb_{0.18}Er_{0.02}F_4@NaYF_4$ (microgauge, solid line, initial mass 5.04 mg) and polystyrene microspheres only (dashed line, initial mass 19.007 mg). Both mass percentages are referenced to the mass remaining after a 20-minute hold at 110°C to account for residual moisture.

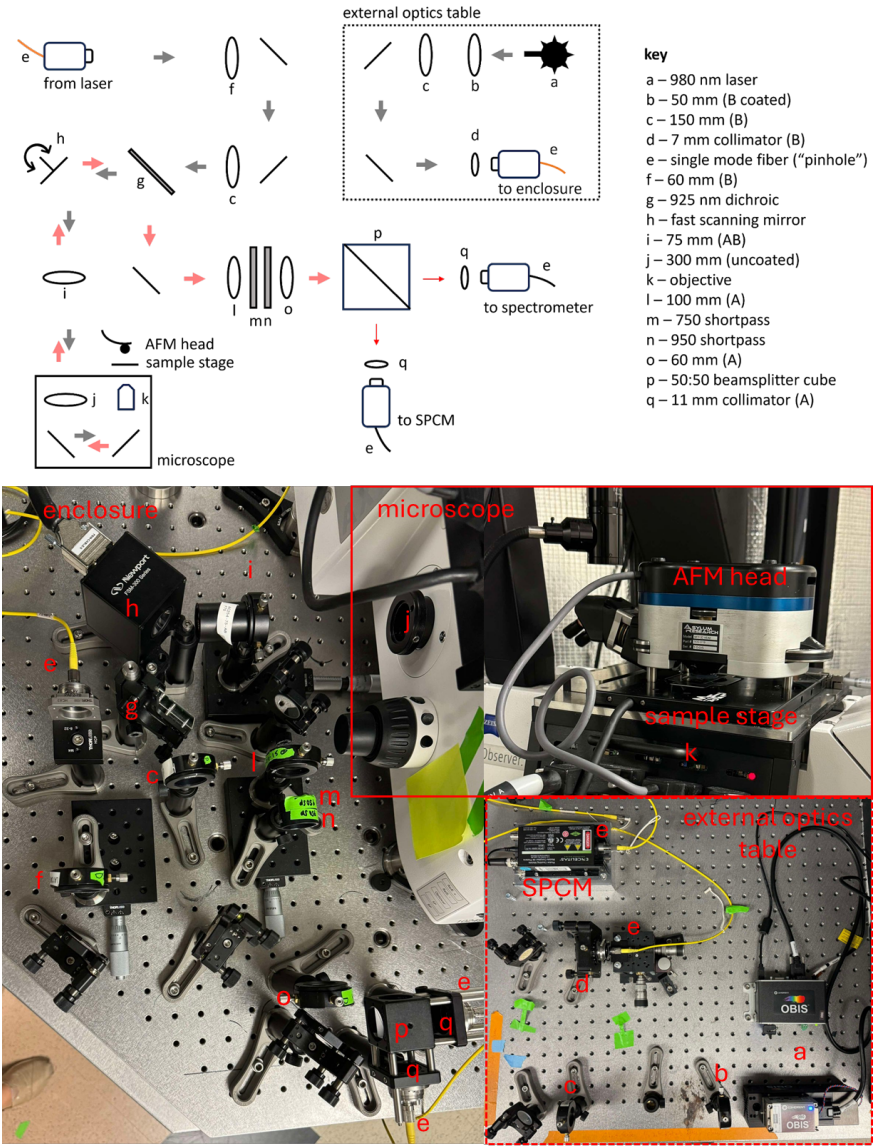

**SI Fig. 4: Confocal optical train**

A schematic diagram (above) and photographs (below) of the confocal optical train. Gray arrows represent the 980 nm excitation path, and red arrows represent the visible (red and green) collection path. All components are housed in a sealed vibration isolation chamber except for those represented inside the dotted outline. Single-mode collection fibers perform the function of the confocal pinhole.

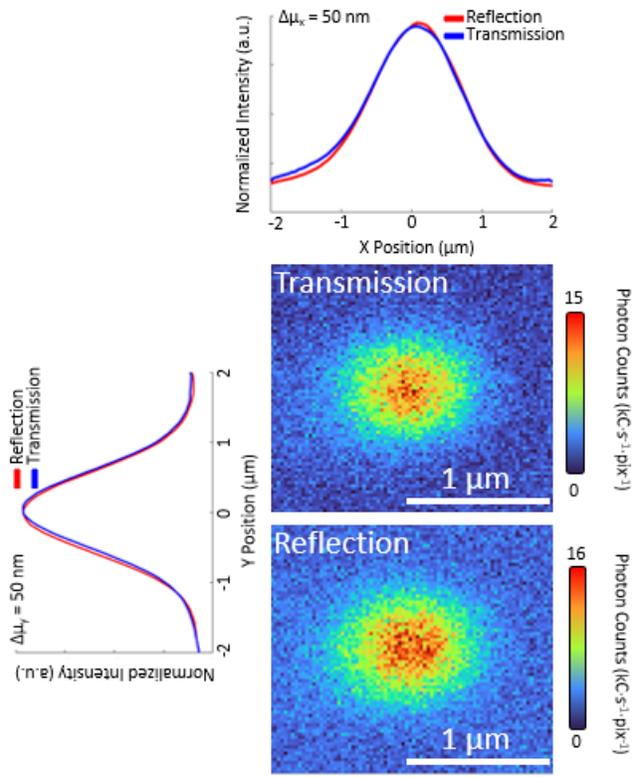

**SI Fig. 5: Co-alignment of two confocal channels**

UCL maps of the same single UCNP taken with the same field of view and focal height showing the co-alignment of two imaging channels downstream of a 50:50 beamsplitter. Images were taken with the same fiber-coupled SPCD, switching between the transmission and reflection fibers. Row and column averages from ten images per channel were fitted to a Gaussian to estimate peak offset in each direction. One pixel offset in both X and Y, representing 50 nm in both directions and ~70 nm total, was measured.

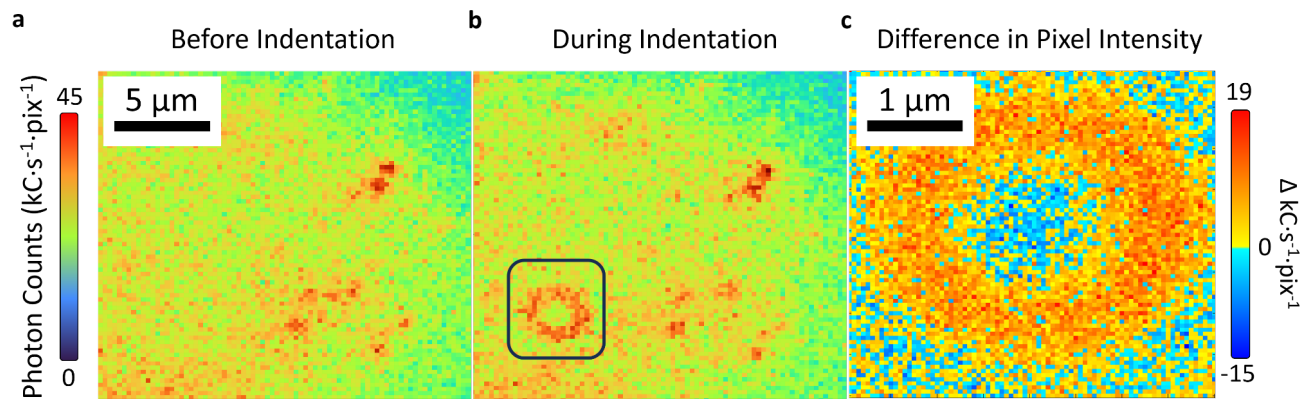

**SI Fig. 6: Colocating the centers of the confocal excitation and the contact region**

**a)** A UCL map taken on the reflection channel of the confocal microscope before indenting the sample with a 10 μm colloidal tip at a trigger force of 1 μN. **b)** During the indentation. Colormaps, scale bars, and fields of view are identical for a and b. **c)** The difference in pixel intensity from the contact region after indentation relative to pixel intensities before indentation.

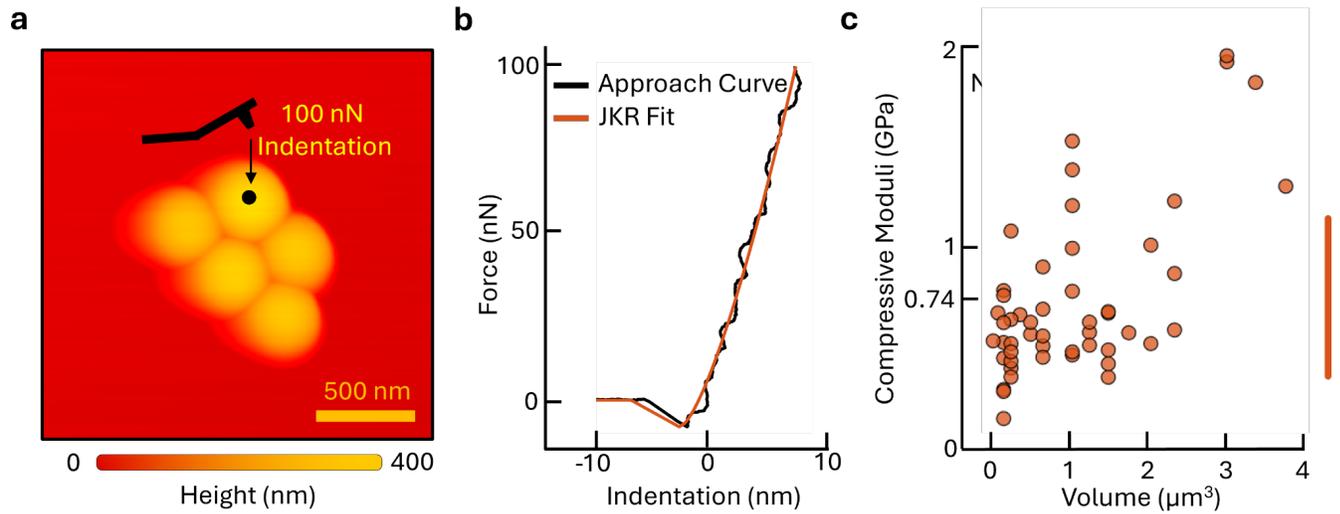

**SI Fig. 7: Method for estimating microgauge compressive modulus**

**a**) An AFM height map taken with a 30 nm radius rounded tip operating in AC mode, showing a cluster of individual microgauges. **b**) A force indentation curve generated from a single microgauge indented to a trigger force of 100 nN (black) and fitted to a JKR model (orange). **c**) All compressive moduli obtained by fitting the approach curves of 50 indented microgauges displayed as a function of particle volume, with the mean (740 MPa) and standard deviation (410 MPa) indicated to the right.

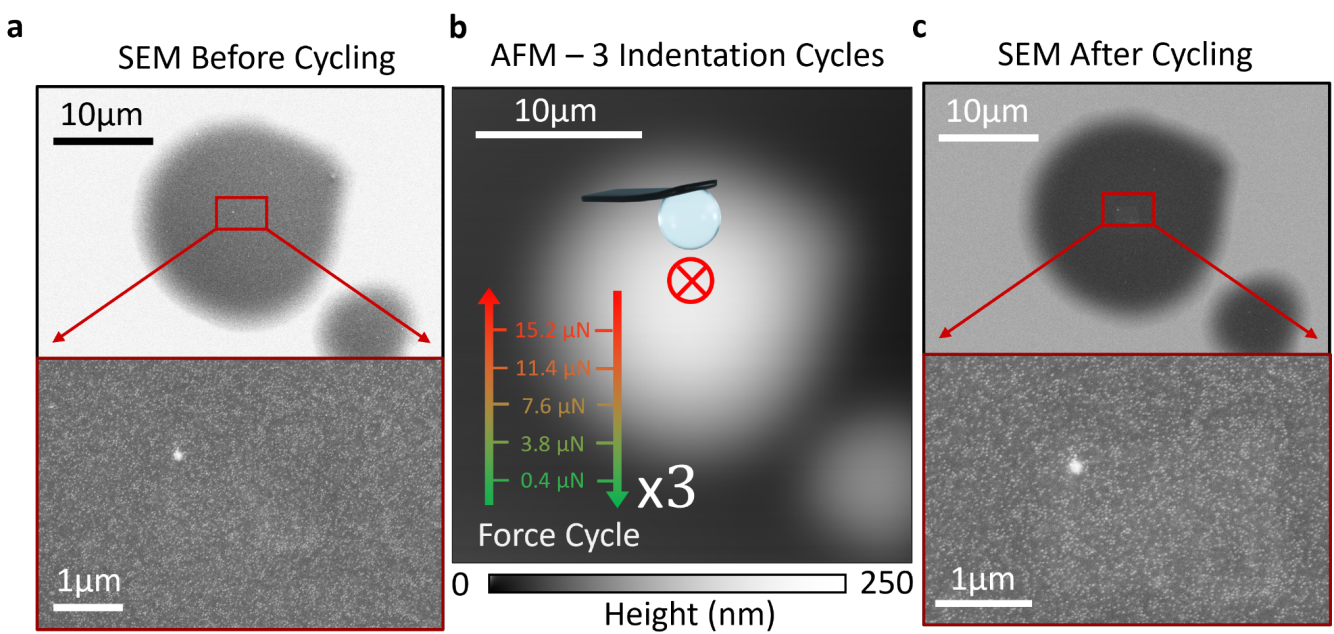

**SI Fig. 8: Microgauge films do not exhibit plastic deformation under the indentation conditions used for optical calibration**

**a)** SEM micrograph of an uncoated patch of microgauge film on silicon taken before indentation, and a higher resolution inset at the indentation region. **b)** A height map of the same patch taken with the 10 μm diameter colloidal tip operating in AC mode. The region indicated with a red X was indented with the same tip operating in contact mode from 0.4 to 15.2 μN and back down again. Each force was held for 30 seconds, and three loading-unloading cycles were performed before withdrawing the tip. **c)** SEM micrograph of the same patch taken after the indentation cycles and a higher resolution inset at the indentation region. The faint rectangular patch of contrast is from beam-induced deposition of organics within the SEM, which occurred during pre-indentation imaging.

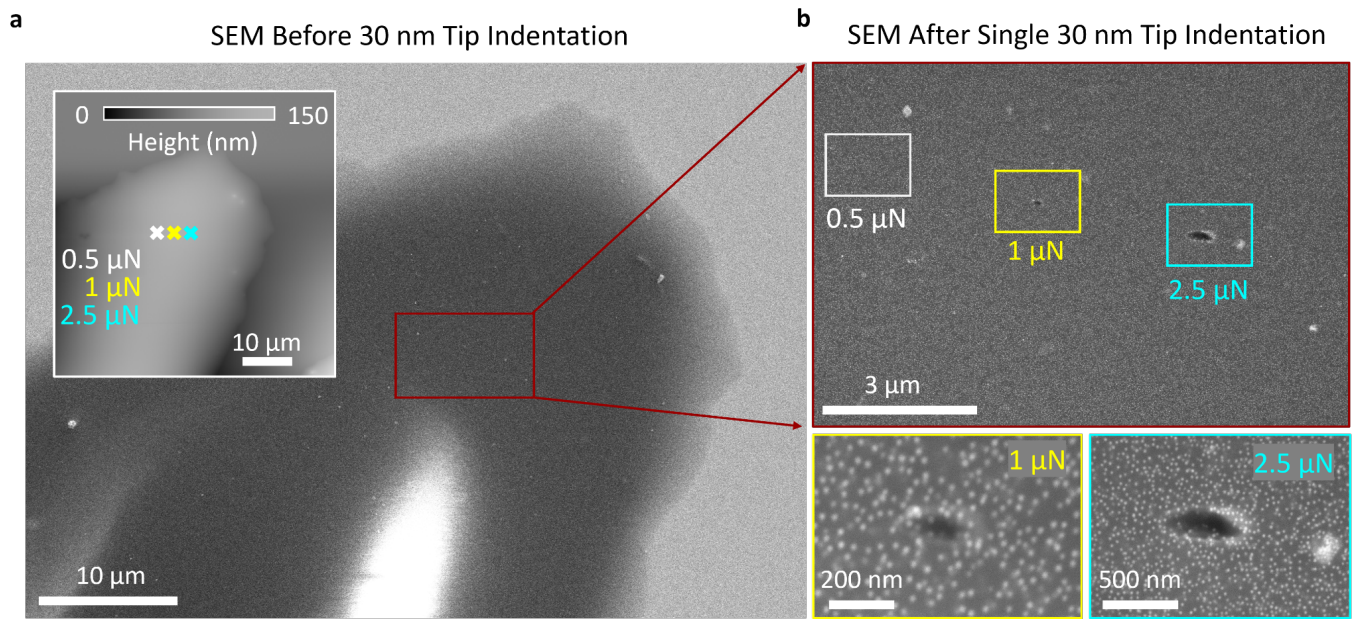

**SI Fig. 9: Microgauge films do exhibit plastic deformation at indentations in excess of the compressive strength of polystyrene.**

**a**) SEM micrograph of a large uncoated patch of microgauge film on silicon before indentation. The inset shows the AFM height map of the same region taken with a 30 nm radius rounded tip operating in AC mode as well as the locations and magnitudes of indentation forces. **b**) SEM micrographs of the same region after indentation, with higher resolution insets showing the film damage occurring at 1 and 2.5 μN (~510 and 700 MPa maximum contact stress, respectively). No scarring was observed in the region around the 0.5 μN indentation, corresponding to ~400 MPa maximum contact stress.

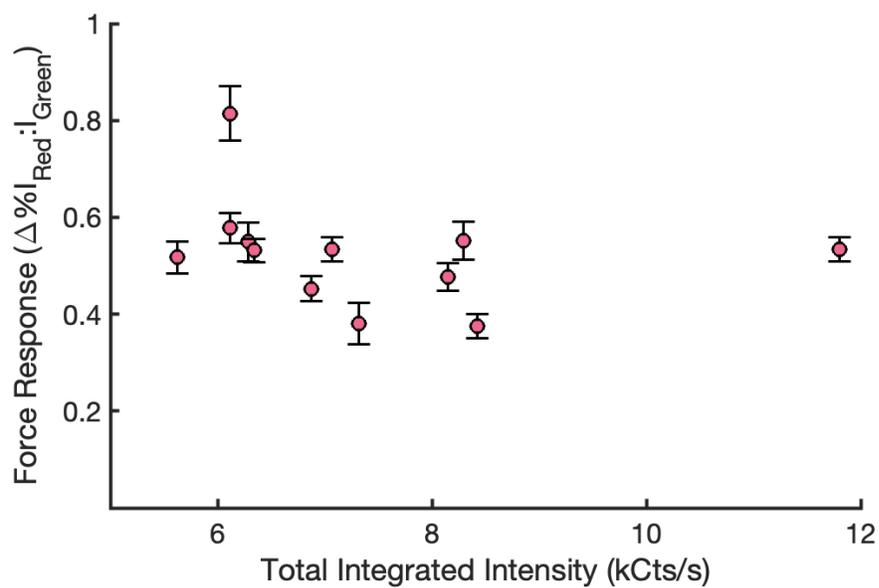

**SI Fig. 10: Correlation between force response and UCNP loading**

The slope of the least squares fit of percent ratiometric change versus force for all 3 cycles of each replicate plotted against the average total integrated intensity in the red and green regions ($I_{Red}+I_{Green}$) for three 90 second integrations taken with the tip engaged at 0.4 µN prior to the indentation cycles. Error bars represent the S.E.

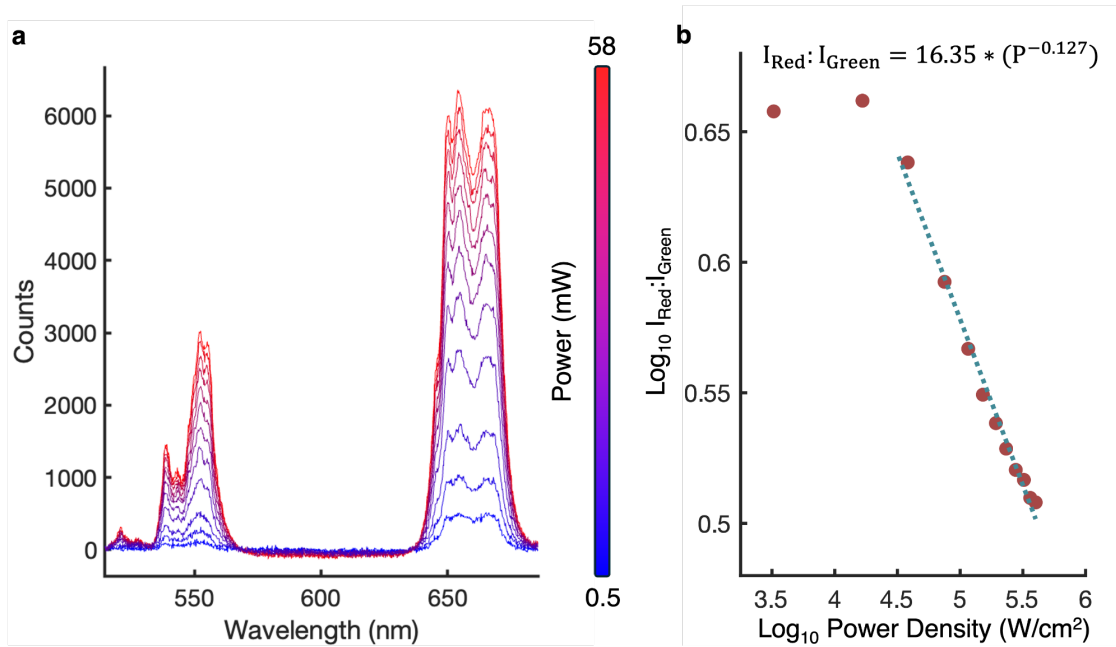

**SI Fig. 11: I$_{Red}$:I$_{Green}$ *vs.* incident laser power**

**a**) Background corrected confocal UCL spectra from a single microgauge taken at different incident powers (0.5 to 58 mW, exiting objective). **b**) A log-log plot of I$_{Red}$:I$_{Green}$ values extracted from the spectra in panel a, fit to a broken power law above 32 kW/cm$^2$.

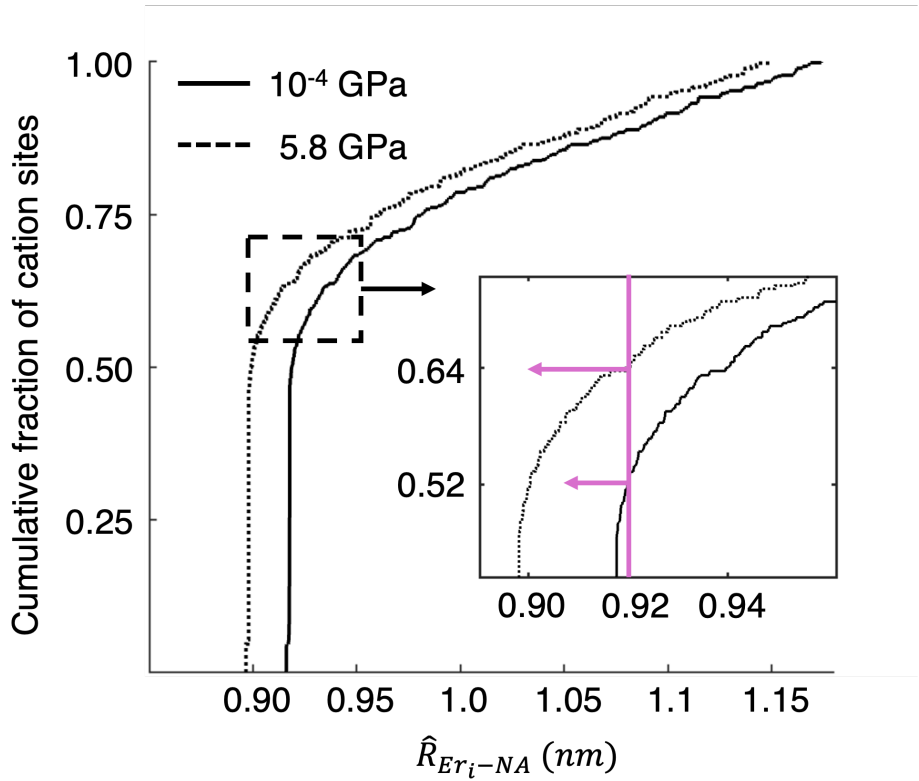

**SI Fig. 12: Expected nearest erbium acceptor distances**

The cumulative distribution function of expected nearest erbium acceptor distances for all cation sites in a 10.9 nm diameter $NaY_{0.8}Yb_{0.18}Er_{0.02}F_4$ core. The calculation was performed at atmospheric pressure (solid) and the maximum pressure employed in the UCL DAC study (5.8 GPa, dashed). The vertical magenta line in the inset indicates the critical Förster radius for cross relaxation determined by [**Rabouw et al. 2018**], and the horizontal magenta arrows are the fraction of sites at each pressure expected to have nearest neighbors closer than that threshold.

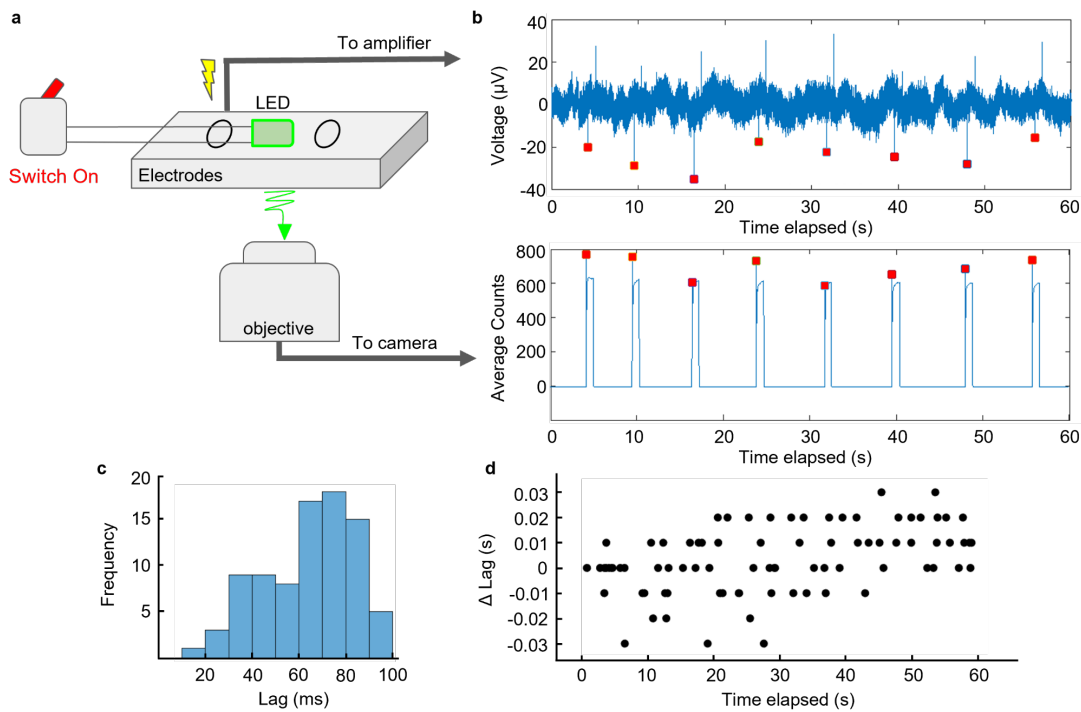

**SI Fig. 13: Lag estimation**

**a)** A schematic diagram of the lag measurement setup wherein a simultaneous electrical and optical impulse are delivered to the two recording streams. The switch was toggled manually between seven and ten times per 60-second trial. **b)** The relative time course of voltage measured on the EPG chip and average pixel brightness measured on the camera from a sample trial. Red squares are used to denote the point on each time course when a change was initially observed. **c)** A histogram of all lags (of the EPG relative to the camera) measured across all ten trials. Each bin has a 10 ms width, and the label represents the upper limit for the bin (inclusive). The minimum lag we observed was 20 ms, and the maximum was 100 ms. **d)** The same lag measurements plotted as their relative change from the initial observed lag within their respective trial. This data represents how synchronicity is affected by data loss in one or both of the recording streams. We note that the units on the x and y axes are both seconds, but the scale of the y axis is drastically exaggerated for clarity.

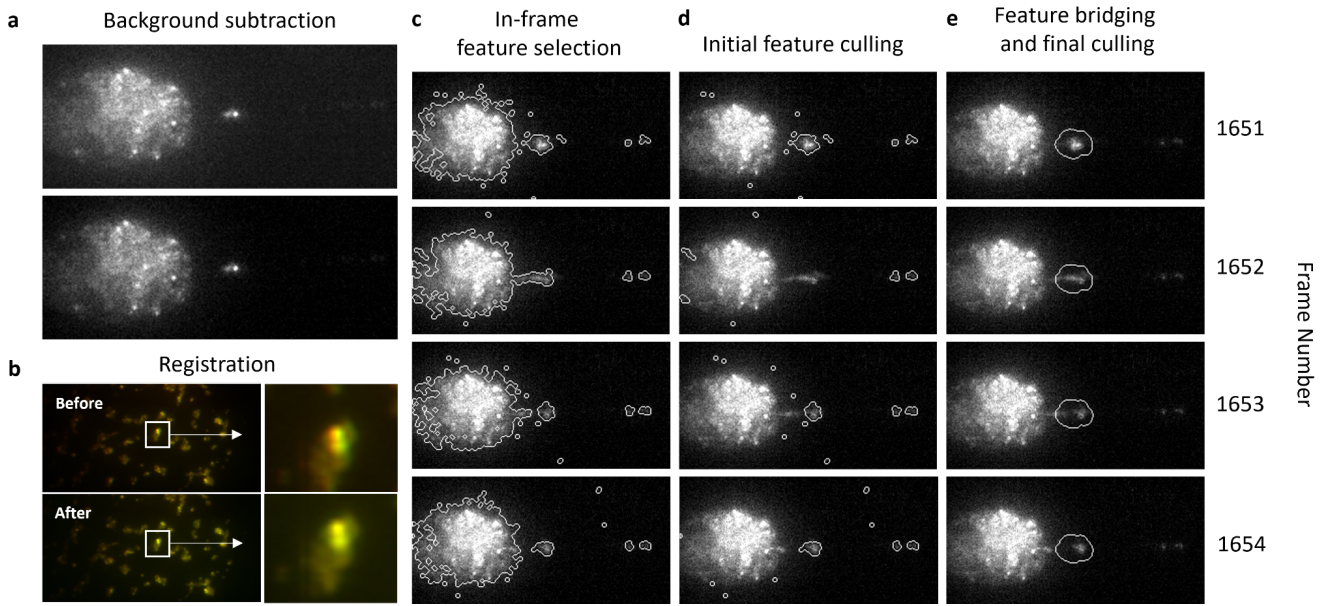

**SI Fig. 14: Optical data processing pipeline**

**a)** One example frame from the red channel of a recorded worm (Fig. 4c.iv) before and after subtracting the background image **b)** An overlay of the red and green channel of the reference image taken after the W-View Gemini lens alignment and before the recording (before), as well as the same image overlay after translational registration (after). **c)** Example frames 1651-1654 from the same recording after the initial binarization and 2D morphological image processing. Outlines represent all contiguous features that survived the thresholding and opening. Notice that in frame 1652, the terminal bulb feature is contiguous with the intestinal feature because material is passing through the VPI into the intestines under the influence of a pump. **d)** The same frames after removing the feature with the largest area, which almost always contains the intestinal signal. Notice that this also erroneously culls the terminal bulb signal in frame 1652. **e)** The same frames after 3D morphological image processing to bridge lost terminal bulb features (ex. 1652) and remove noise. The outline depicted in frame 1652 was thus interpreted from the spatial information of the features in a set of neighboring frames. If, as in this case, the anterior pharynx features survived this process, they were removed manually. The segmentation output for this 3000 frame recording is presented in full in SI Video 6.

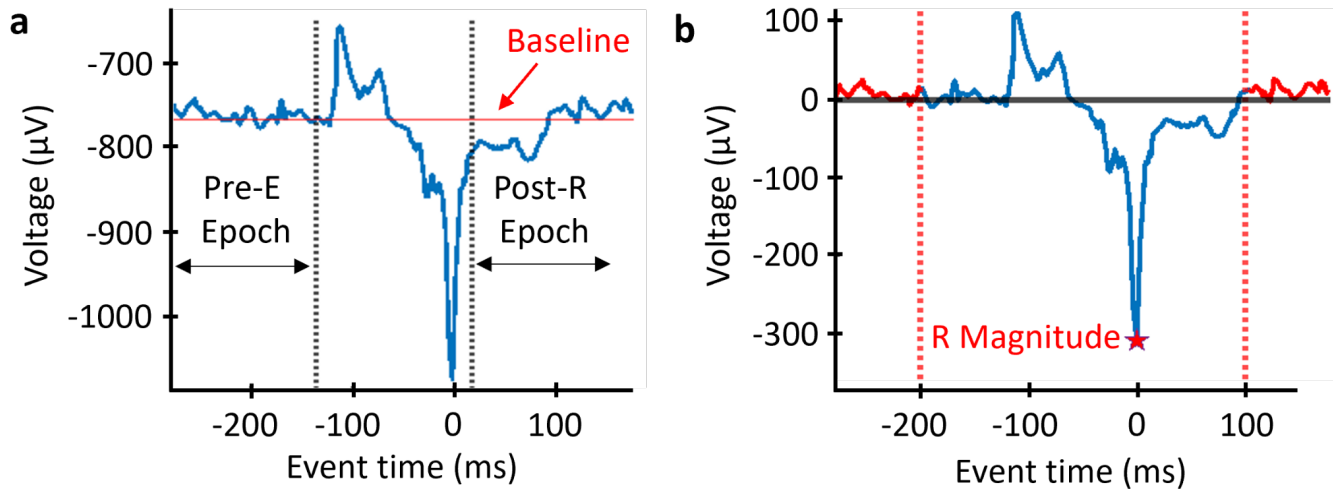

**SI Fig. 15: Electrical baseline subtraction**

An illustration of the event window before time normalization as well as the voltage time course from a sample pump **a**) before and **b**) after DC offset. This baseline correction is calculated from the mean of the Pre-E and Post-R Epochs. Once corrected, the RMS of the event noise (red regions) is calculated and compared to the magnitude of the R peak. As illustrated, in the absence of low-frequency interference, this value should be well below the 20% threshold used for quality control. See SI Fig. 16e for examples of the voltage-time courses that do not pass quality control.

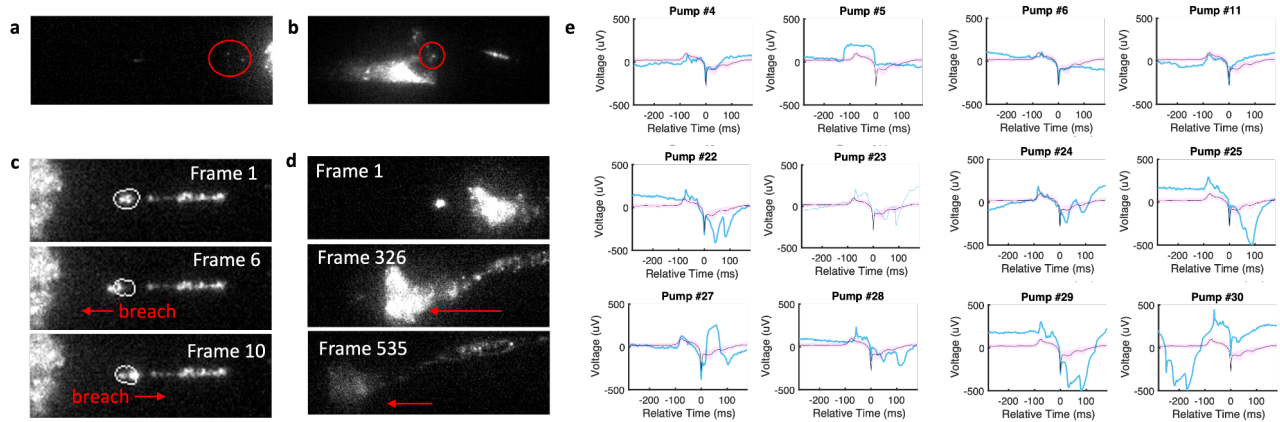

**SI Fig. 16: Exclusion criteria**

An illustration of several conditions that resulted in optical data set exclusion either **a,b,d** before or **c**, after segmentation, as well as the individual events that were excluded based on the electrical data set within a sample data set (Fig. 4c.iv) that did pass quality control. **a)** A frame illustrating a lack of sufficient terminal bulb accumulation (red circle), as well as a lack of accumulation in the anterior pharynx to eventually replenish it. **b)** A frame illustrating the difficulty of isolating terminal bulb signal (red circle) in the morphological image processing step when the terminal bulb is persistently too close to the intestines. **c)** A representative set of three frames from a segmentable data set that was given an average, blinded segmentation score of 1.3. As illustrated, this data set exhibited frequent segment "breaches" during pumps, which resulted in the measured ratiometric changes not being representative of those in the terminal bulb as a whole. Such breaches usually occur in recordings where the segment needs to be made small enough to avoid capturing luminescence from microgauges accumulated in the anterior isthmus. **d)** Three frames illustrating the progressive backward drift of a worm. Aside from insufficient accumulation, this was the most commonly observed condition that resulted in data set exclusion. **e)** All 12 events (blue) in the order they were observed that were excluded from Fig 4c.iv because of an RMS value that exceeded 20% of the R magnitude. "Relative time" is used synonymously with "event time." The average (black) and standard deviation (magenta) of the 18 surviving events are overlain for comparison.

## Supplementary Citations

**Supplementary Videos**

Available at https://doi.org/10.25740/ff923hb3417

**Supplementary Video 1 - Microgauge cross sections**: Internal cross sections of a microgauge taken in a focused ion beam scanning electron microscope after platinum deposition (outer white layer). The polystyrene appears black and the UCNPs embedded inside appear as white dots.

**Supplementary Video 2 - Microgauge transport under pharyngeal pumping**: A video (66 fps) of microgauge transport under the influence of pharyngeal pumping in the terminal bulb. Dual illumination from a lamp (above) and a 980 nm laser (below) was used to make both the pharynx and the microgauges visible. Microgauge UCL is false colored in cyan. Scale bar is 50 μm.

**Supplementary Video 3 - Microgauge ingestion in a freely crawling worm**: A worm crawling on a lawn of microgauges and *E. coli* ingesting material and transporting it through the pharyngeal lumen to the intestines. The pharynx is facing in the direction of movement. Video has been slowed down 2.5 times. Scale bar, 250 μm.

**Supplementary Video 4 - Microgauge defecation in a freely crawling worm**: A worm crawling on a lawn of microgauges and *E. coli* undergoing a single defecation motor program cycle to expel material. Material can be seen moving anteriorly under the influence of a posterior body contraction then posteriorly under the influence of an anterior body contraction before the sphincter muscle rapidly and briefly contracts to expel material. The pharynx is facing left for the duration of the video. Video has been slowed down 2.5 times. Scale bar is 250 μm.

**Supplementary Video 5 - Microgauge defecation in a channel immobilized worm**: A UCL video (10x, 50 fps, 2s duration) of microgauge transport in the posterior intestinal lumen under the action of the defecation motor program. The approximate onset of posterior body contraction (pBoc), anterior body contraction (aBoc) and expulsion (Exp) events are labelled in the appropriate frames. Scale bar is 50 μm.

**Supplementary Video 6 - Full segmentation example**: A red UCL video (50x, 50 fps, 60s duration) of microgauges during pharyngeal pumping. The outline indicates the bounds of the pixel segment used to calculate the optical time series. This is the same dataset used to calculate the event-triggered averages shown in Fig. 4c.iv. We note that, out of the 30 pumps identified from the corresponding EPG, only 12 had noise levels low enough to meet our quality control standards (see Supplementary Fig. 16 for those that did not). Scale bar is 20 μm.